\documentclass[aps,pre,twocolumn,superscriptaddress,showpacs]{revtex4}

\usepackage{graphicx}
\usepackage{subfigure}
\usepackage{amssymb,amsmath} 

\usepackage{color}
\definecolor{orange}{rgb}{1.0,0.6,0.4}
\definecolor{yellow}{rgb}{1.0,1.0,0.0}
\definecolor{green}{rgb}{0.4,1.0,0.4}
\definecolor{blue}{rgb}{0.5,0.4,1.0}
\definecolor{red}{rgb}{1.0,0.0,0.0}
\usepackage[normalem]{ulem}
\makeatletter
\def\marker#1{\bgroup
\color{#1}%
 \def\UL@start{\setbox\UL@box\hbox\bgroup\everyhbox{\UL@hrest}%
  \let\UL@start\@empty \def\UL@unegroup{\bgroup\bgroup}\let\UL@leadtype\@empty
  \bgroup \color{black}%
  \kern-3sp\kern3sp 
  \if@ignore \global\@ignorefalse \ignorespaces \fi}
  \def\UL@leadtype{%
     \leaders \hrule \@height\ht\strutbox \@depth\dp\strutbox }%
  \ULon}
\makeatother

  \usepackage{ifpdf}

\begin{document}


\title{An Application of a Self-consistent Mean-field Theoretical Model
to POPC-PSM-Cholesterol Bilayers}

\author{Paul W. Tumaneng}
\affiliation{Department of Biological, Chemical and Physical Sciences and 
Center for the Molecular Study of Condensed Soft Matter, Illinois 
Institute of Technology,  Chicago, IL 60616, USA}

\author{Sagar A. Pandit}
\affiliation{Department of Physics, University of South Florida, Tampa, FL 
33620, USA}

\author{Guijun Zhao}
\affiliation{Department of Biological, Chemical and Physical Sciences and 
Center for the Molecular Study of Condensed Soft Matter, Illinois 
Institute of Technology, Chicago, IL 60616, USA}

\author{H.L. Scott}
\affiliation{Department of Biological, Chemical and Physical Sciences and 
Center for the Molecular Study of Condensed Soft Matter, Illinois 
Institute of Technology, Chicago, IL 60616, USA}


\date{\today}

\begin{abstract}
The connection between membrane inhomogeneity and the structural basis of lipid rafts has sparked interest in the lateral organization of model lipid bilayers of two and three components. ÊIn an effort to investigate anisotropic lipid distribution in mixed bilayers, a self-consistent mean-field theoretical model is applied to palmitoyloleoylphosphatidylcholine (POPC) - palmitoyl sphingomyelin (PSM) - Cholesterol mixtures.  The compositional dependence of lateral organization in these mixtures is mapped onto a ternary plot.  The model utilizes molecular dynamics simulations to estimate interaction parameters and to construct chain conformation libraries.Ê We find that at some concentration ratios the bilayers separate spatially into regions of higher and lower chain order coinciding with areas enriched with PSM and POPC respectively.  To examine the effect of the asymmetric chain structure of POPC on bilayer lateral inhomogeneity, we consider POPC-POPC interactions with and without angular dependence.Ê   Results are compared with experimental data and with results from a similar model for mixtures of dioleoylphosphatidylcholine (DOPC) , steroyl sphingomyelin, and Cholesterol.
\end{abstract}


\maketitle

\section{\label{intro}Introduction}
The membranes of cells are composed of a variety of amphipathic lipids that spontaneously
form bilayers in the presence of water.  In nature, the lipids of typical plasma membranes
vary by head-group, chain length, and chain saturation.  The exact composition, as well
as external factors such as temperature and pressure, strongly affect the properties of the
membrane, so many experiments have been aimed at characterizing model lipid 
bilayers of different compositions as a function of these external factors~\cite{veatch,haluska,
kahya,veatch3,stottrup,aittoniemi, zhang,dealmeida2, lindblom, zhao, collins, heberle, moore,
zheng,zhao2, filippov1, filippov2,cicuta,castro, bakht, frazier, ursell, veatch4,
pokorny, bunge, halling,dealmeida,tsamaloukas}.
In particular, the lateral organization of ternary-component
mixtures has been of recent interest because      they contain essential ingredients for the
formation of lipid rafts, under the right conditions.  
Lipid rafts have been the subject of a number of reviews~\cite{lingwood, simons, edidin, jacobson,
hancock, pike1, pike2} where they are described as  nanometer-scale sphingolipid and 
cholesterol-enriched domains whose importance ranges from signal transduction to
organization of bioactivity in cell membranes.

Bilayers composed of a single lipid type can undergo a main-chain phase transition between
a liquid disordered phase, $l_\alpha$, and a gel phase, $s_0$, as a phase transition 
temperature, $T_m$, unique to the lipid type, is traversed.  The $l_\alpha$ phase is characterized
by highly mobile lipids with disordered chains, while in the $s_0$ state the lipids are much
less mobile and their chains are more ordered.  

A side-by-side comparison of two bilayers of nearly identical composition, with only a
minor difference such as the presence of a double bond in one of the lipid chains, can 
underscore the complex nature of real cell membranes.  For example, 
dioleoylphosphatidylcholine (DOPC) has two mono-unsaturated 18-carbon fatty acid 
chains, while in palmitoyloleoylphosphatidylcholine (POPC) one of those is replaced
by a 16-carbon saturated chain.  
As a result, the DOPC main-chain phase transition temperature is lower than 
that of the POPC temperature by 26$^\circ C$. 
Additionally, the area per lipid of POPC bilayers,  
65 \AA{}$^2$~\cite{yeagle}, is significantly lower than that of DOPC bilayers, 
72 \AA{}$^2$~\cite{lantzsch}.  The presence or absence of the double bond clearly plays 
a role in how well neighboring lipids pack with each other, which in turn determines
the physical properties of the bilayer.

Binary mixtures of low $T_m$ phosphatidylcholines and high $T_m$ phosphatidylcholines
or sphingomyelin (SM) can exhibit phase separation between the $l_\alpha$ phase and $s_0$ 
phase at temperatures intermediate to their respective melting 
points~\cite{dealmeida,veatch,veatch2,korlach}.  
The phase coexistence depends upon the relative
concentrations of each lipid type, and can be observed directly in fluorescence microscopy
experiments~\cite{veatch,veatch2}.

Cholesterol (CHOL), another major membrane lipid, has a strong and complex effect on bilayer
properties.  CHOL has a tendency to induce order in the chains of $l_\alpha$-phase
lipid bilayers and increase mobility of $s_0$-phase lipid bilayers.  When the CHOL concentration is above about 15\% the resulting structural state 
is sometimes called the liquid ordered phase, $l_0$, and is characterized by highly ordered, yet
mobile chains.  The phase behavior of binary mixtures of CHOL and other membrane 
lipids has been studied extensively~\cite{vist,mcmullen,dealmeida,mateo,scmft1, scmft2}.
Pan \emph{et al}.~\cite{pan} find that CHOL-lipid interactions depend 
on the number of saturated chains present, affecting physical properties such as
bending modulus, area per lipid, and order parameter.
Other studies~\cite{ramstedt,tsamaloukas,halling,aittoniemi} indicate that CHOL has a higher
affinity for one lipid type over another, for example dipalmitoyl phosphatidylcholine (DPPC) or di-saturated sphingomyelin (SM) over POPC.   CHOL shows a 
unique level of specificity in inducing order because the two ``faces" of CHOL are 
structurally distinct.  Here, ``face'' refers to the sides of the flat fused ring structure of CHOL,
one of which has protruding methyl groups (rough $\beta$ face), while the 
other does not (smooth $\alpha$ face).  Pandit \emph{et al}.~\cite{langmuir} utilize atomic-level 
simulations of lipid-CHOL mixtures to reveal that the effective molecular area of CHOL is smaller in POPC than in DOPC and even in DPPC. They attribute this effect to a combination of the anisotropic chain structure of POPC and the anisotropic faces of CHOL.

This differential behavior of CHOL in lipids must have a strong effect on the lateral organization of ternary component mixtures
consisting of a high $T_m$ lipid, a low $T_m$ lipid, and CHOL.
Experiments~\cite{veatch,frazier, dealmeida, bunge, pokorny, baumgart, morales} 
consistently indicate that these ternary component 
mixtures have a complex lateral structure that is highly dependent 
on the relative abundance of components and external parameters such as temperature
and pressure.  In particular, at certain concentration ratios, the liquid phases $l_\alpha$ and 
$l_0$ are known to coexist and are the basis for the concept of ``lipid rafts'' which 
are believed to exist in biological plasma membranes~\cite{jacobson, marsh,pike1,pike2}, 
and are implicated in a number of biologically important cellular functions 
(See e.g.~\cite{hancock,simons}).

A number of experimental treatments of ternary mixtures have been published~\cite{haluska,
kahya,veatch3,stottrup,aittoniemi, zhang,dealmeida2, lindblom, zhao, collins, heberle, moore,
 zheng,zhao2, filippov1, filippov2,cicuta,castro, bakht, frazier, ursell, veatch4, lichtenberg}, 
but the details on the formation of rafts in POPC-SM-CHOL mixtures remains a subject 
of debate.  A number of experimental studies have treated specifically this system~\cite{marsh,
pokorny, bunge, halling,dealmeida,frazier,tsamaloukas}. Veatch and  Keller~\cite{veatch}
report liquid-liquid phase coexistence in giant unilamellar vesicles using fluorescence
microscopy.  Zhao \emph{et al}.~\cite{zhao2}  do not observe liquid-liquid phase separation in POPC-SM-Chol mixtures
at the micron level.  They do observe $l_d$ and $s_0$ phase coexistence and 
suggest that domains may exist in smaller sub-micron 
clusters.  Overall, current experimental studies underscore a need to carefully probe  
subtle atomic level effects in the POPC-SM-CHOL ternary system.

Several recent theoretical models have been published that describe
bilayer phase behavior in three-component systems.
Putzel \emph{et al.}~\cite{putzel1, putzel2} have proposed a phenomenological model
intended to elucidate the mechanism for phase separation in ternary lipid systems
by examining phenomenological free energy functions for mixtures 
consisting of saturated-chain lipids, unsaturated-chain lipids and CHOL.
They were able to reproduce ternary phase diagrams that
are consistent with those proposed by experiment.
Elliot \emph{et al.}~\cite{elliott, elliot2} have proposed a self-consistent 
mean-field theoretical (SCMFT) model of
a three-component system that treats chain interactions at an atomic level.
The model employs equilibrium statistical mechanical analysis
to construct ternary diagrams with phase boundaries whose presence and location
are modulated by interaction parameters.
We recently published a model for ternary lipid mixtures that we
applied specifically to DOPC-Steroyl sphingomyelin (SSM)-CHOL bilayers~\cite{jcp}.  The model projected  three-dimensional
lipid bilayer leaflets onto two-dimensional fields of chain order over which CHOL
could diffuse.  This model, based on a combination of equilibrium statistical mechanics and Langevin plus Cahn-Hilliard dynamics, described temporal organization of the mixed bilayer. Results were displayed on a triangle diagram that agreed well with experiment for  DOPC-SM-CHOL mixtures.

In this paper we present a model of ternary-component lipid bilayers that builds on our previous
modeling work~\cite{jcp, scmft1, scmft2, scmft3} to include lipids
with non-identical chains.  This model differs from other computational and theoretical 
models~\cite{putzel1,
 putzel2, shi, idema, elliott, elliot2, fattal, ben-shaul, gruen, muller, mcconnell, risselada} in
 that it utilizes atomistic molecular dynamics (MD)-generated data as input to a self-consistent 
 mean-field theoretical (SCMFT) model
 that is used to characterize the structure and temporal evolution of ternary phase diagrams.  We apply the model to
POPC - Palmitoyl sphingomyelin (PSM) - CHOL mixtures, but the methodology can be
 generalized to include any two-chain lipid ternary mixture.  
To account for the asymmetric chain structure of the POPC molecule,
an orientational component for POPC is added to the interaction energy function.
 We show below that the effect of including orientational dependence is to slightly amplify the separation of
 lipids into regions rich in PSM separated from regions rich in POPC.  Comparison with our SCMFT simulations of DOPC-SSM-CHOL \cite{jcp} reveals that the degree of lateral organization in POPC-PSM-CHOL mixtures is reduced, and differences between order parameters in separated domains is more subtle.
In the next section, the POPC-PSM-CHOL ternary system is described.  Subsequent sections present our results
 and discussion.

\section{\label{model}Theoretical Model}

In our SCMFT model,~\cite{jcp, scmft1, scmft2, scmft3}, a three-dimensional lipid bilayer
leaflet is cast as a two-dimensional field of weighted, chain-averaged order parameters:
\begin{equation}
  \label{op}
  s(\overrightarrow{r}) = -\frac{n_{tr}}{n_s} \sum_{m=1}^{n_s}
  (\frac{3}{2} cos^2 \beta_{m} - \frac{1}{2} ) / n_s,
\end{equation}
where the weighting fraction $\frac{n_{tr}}{n_s}$ represents the fraction of dihedrals in a
trans configuration ($n_{tr}$) along a single chain to the number of dihedrals along
that chain ($n_s$).  $\beta_{m}$ is the angle between the 
C-H bond vector and the bilayer normal
for carbon m for the chain at the position $\overrightarrow{r}$.
CHOL molecules are treated as 
two-dimensional `rods' that are free to diffuse through the order parameter field.
Coupled to this order parameter field is a composition field whose purpose is to identify
the concentration of lipid type at any point.  Both fields are discretized onto an underlying lattice for computation. The overall model methodologies are as follows: Langevin dynamics are used to propagate CHOL molecules over the two dimensional fields, and to locally reorient POPC molecules. Cahn-Hilliard dynamics are used to model the evolution of lateral compositional order in the composition field. Mean field statistical mechanics is employed to recalculate the order parameter field after each dynamical timestep. These methods are described in the following subsections.

\subsection{The POPC-PSM-CHOL Ternary System}

\begin{figure}[htb]
\centering
   \includegraphics[width=85mm]{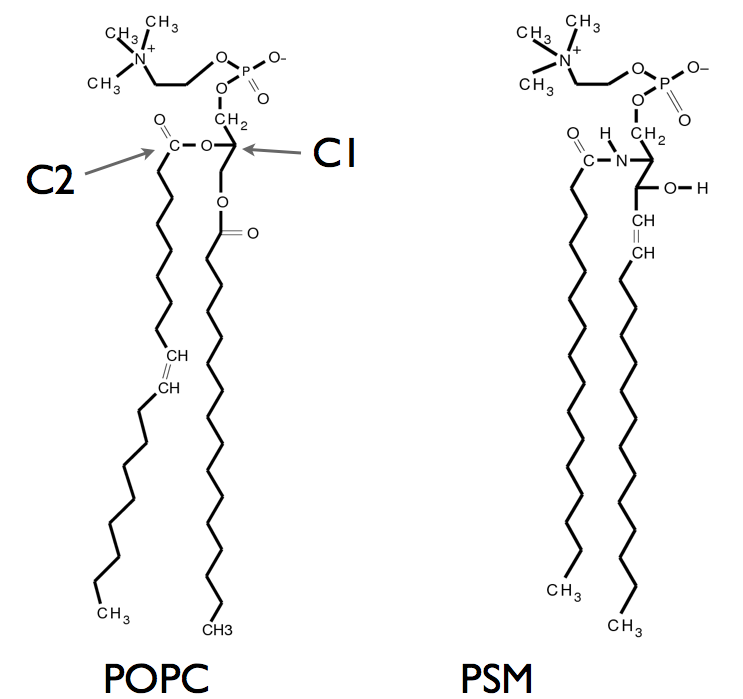}
      \caption{The structures of POPC and PSM.  The labelled arrows
      indicate specific atoms used to define POPC molecular orientation. }
      \label{popc_psm}
\end{figure}

The structures of POPC and PSM are shown in Fig \ref{popc_psm}.  POPC has two
non-identical hydrocarbon chains; one saturated palmitoyl (16:0) chain and one unsaturated
oleoyl (18:1) chain.  PSM ((16:0)SM) consists of a palmitoyl chain attached to a sphingosine
backbone.  PSM has a higher main-chain phase transition temperature 
(40$^\circ C$)~\cite{maulik} than POPC (-3$^\circ C$).  For this model, we focus on
the temperature range intermediate to the two
phase transition temperatures, namely 30$^{\circ}$ C.

In contrast to our previous implementation of the SCMFT model~\cite{jcp} 
for DOPC-SSM-CHOL mixtures, we now
consider a  case where the low melting point lipid, POPC,  has
nonequivalent acyl chains.
To address the nonequivalence of POPC chains,  we consider the effect of adding an orientation degree of freedom to each POPC molecule, as we describe below.
Modifications to the previous model \cite{jcp} include redefining
the order parameter, making adjustments to the Hamiltonian
and mean-field free energy, and modifying the method of updating time steps.  
The modifications are summarized in the following subsections.

\subsection{Order Parameter Field and Angle Field}

The SCMFT model~\cite{jcp,scmft1,scmft2,scmft3} we previously applied to binary and ternary mixtures
of lipids with two chains of identical structure projected single chain order parameters onto
a two dimensional lattice.  Lattice grid sites were spaced at a distance calculated
from MD simulations as the average distance between two neighboring chains.
To incorporate the asymmetric structure of POPC molecules,
we now define  a `whole-molecule' 
order parameter for each site as an average:
\begin{equation} s_{mol} = \frac{1}{2} (s_{sn1} + s_{sn2})
\label{eq:two-chain}
 \end{equation}
where $s_{sn1}$ and $s_{sn2}$ are the order parameters for the two chains on each molecule, as defined in Eq \ref{op}.
In this work, we take the lattice spacing to be the average nearest neighbor center of mass distance between molecules.
Although the two chains in PSM do not differ appreciably in order parameter value,
PSM moleular order parameters are also calculated using Eq \ref{eq:two-chain} to keep the model internally consistent.

\begin{figure}[htb]
\centering
   \includegraphics[width=85mm]{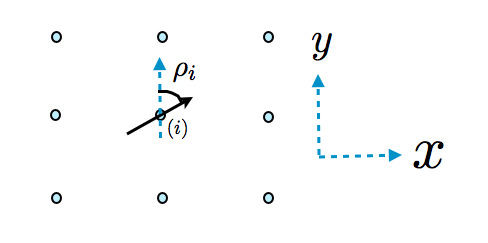}
      \caption{The angle $\rho$ for a POPC molecule at site $i$.  
      The black arrow represents the orientation vector
      of the lipid at site $i$, pointing from the saturated chain of POPC towards the unsaturated chain.  
      We measure the angle $\rho$ with respect to the $y$ direction of the $x-y$ axis of the lattice.}
      \label{orientation}
\end{figure}

In order to incorporate the asymmetric structure of POPC molecules, 
a molecular orientation variable, $\rho$, is introduced and is associated with each
lattice point that contains all or part of a POPC in the local concentration field. For a POPC molecule, $\rho$ is calculated as the angle made between a vector
which points from the atom labeled `c1' towards the atom labeled `c2' in Fig \ref{popc_psm}
and a fixed direction in the plane of the bilayer (Fig \ref{orientation})

The angle field is assumed to evolve dynamically 
according to a Langevin equation:
\begin{equation}\label{angle_langevin}
\frac{\partial \rho_i}{\partial t} = 
- M_{\rho} \frac{\partial F}{\partial \rho_i}
+ \xi_i .
\end{equation}
$M_{\rho}$ is the mobility of the 
angle as it diffuses azimthally at the site $i$ and is related to the diffusion constant
by $M_{\rho} =  D_{\rho}/k_b T$.  $F$ is the free energy, 
shown below, and $\xi_i$ is a stochastic noise component that satisfies 
fluctuation and dissipation relations.   

\begin{figure}[htb]
\centering
   \includegraphics[width=85mm]{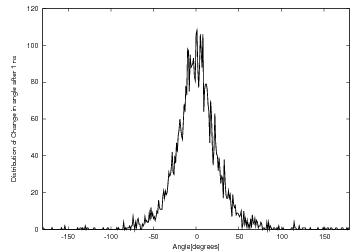}
      \caption{Distribution of the change in angle after 1 ns for MD-simulated POPC molecules.}
      \label{theta_diff}
\end{figure}

The diffusion constant $D_{\rho} $ is found by performing statistical analysis on measurements 
of the angle $\rho$ calculated directly from molecular dynamics simulations.
This angle is recalculated after a nanosecond of simulation has passed. 
Fig \ref{theta_diff} shows a distribution plot of trajectory-averaged changes in the orientation 
angle after one nanosecond intervals
for all of the POPC molecules in a MD simulation (details of the simulation are discussed below), 

\begin{figure}[htb]
\centering
   \includegraphics[width=85mm]{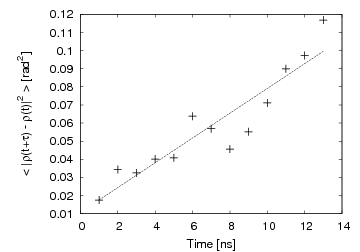}
      \caption{Average square change in the POPC orientation angle as a function of time.  
      The slope of the best fit line (dotted line) yields $2 D_{\rho}$.}
      \label{bfl}
\end{figure}

The gaussian-like shape of the change in angle over time allows us employ an Einstein relation to extract the orientational diffusion constant from MD simulations:
\begin{equation}
\left< |\rho(t+\tau) - \rho(t)|^2 \right> = 2 D_{\rho} \tau
\end{equation}
over a short time $\tau$ giving us a diffusion constant in units of $radians^2/ns$.
A best-fit line of a plot of $\left< |\rho(t+\tau) - \rho(t)|^2 \right> $ vs. $\tau$,
shown in Fig \ref{bfl}, yields twice the diffusion constant.  The numerical value
of $D_{\rho}$ and other parameters used in this work are given in table \ref{tpopc:}.

\subsection{Concentration Field}

To model two different types of lipids in our mixtures, we include a concentration field,
$\psi_i(t) = c_{psm,i}(t) - c_{popc,i}(t)$ which describes the compositional makeup of site i at time t.
As in our earlier work, ~\cite{jcp},
$c_{psm,i} (t) $ is the concentration of PSM at site $i$, and $c_{popc,i} (t) $ is the concentration
of POPC at site $i$, both of which range from 0 to 1.
Therefore, $\psi_i (t) $ varies continuously from $-1$ to $+1$ with the $-1$ representing
pure POPC and $+1$ representing pure PSM.
Locally, $\psi_i (t) $ may vary but the total number of lipids present in the system
remains constant and so the sum over all sites, $\sum_i \psi_i (t) $, is fixed at each time.
The concentration field evolves at each lattice point through the Cahn-Hilliard equation \cite{chaiken}:
 \begin{equation}
 \label{ch}
 \frac{\partial\psi_i}{\partial t} = 
 \Gamma \nabla^2 \frac{\partial F}{\partial \psi_i}+\gamma_i.\end{equation}
  $F$ is the free energy and $\gamma_i$ is a conserved stochastic noise component.
$\Gamma$ is the lipid mobility and is related 
to the lipid diffusion constant
by $\Gamma =  D_{lipid}/k_b T$.   The value of $D_{lipid}$ is unchanged from earlier work, and is given in Table 1.
Eq \ref{ch} ensures that, while
 local lipid concentration field values may vary in time in response to thermal fluctuations
 and tendency towards free energy minimization, the total lipid concentrations
 in the system are conserved. 
 
\subsection{Hamiltonian}
 
The system Hamiltonian contributions arise from lipid-lipid
chain interactions, lipid chain-CHOL interactions, and CHOL-CHOL interactions:
\begin{equation} \label{pham} H_{tot} = H_{lip-lip} + H_{lip-CHOL} +
  H_{CHOL-CHOL}. \end{equation}
The Hamiltonian couples the order parameter and concentration fields to each other
and to the overlying CHOLs.  Due to the anisotropic nature of the POPC molecules, the first term in Eq \ref{pham} differs
from our previous model~\cite{jcp} and is explained in detail below.  The second and third terms in 
Eq \ref{pham} have not been altered from \cite{jcp}, but are also briefly summarized below.
    
\subsubsection{Lipid-Lipid, Interaction Term,  $H_{lip-lip}$}
Lipid-lipid interactions are calculated following the model first proposed by Mar\u{c}elja~\cite{marcelja},
defined in terms of  order-parameter based pairwise interactions between chains.
The model was originally applied to DPPC-CHOL mixtures~\cite{scmft1,scmft2,jcp}, and produced heat capacities and a phase diagram that agreed quite well with experiment~\cite{vist}.  In this application, we consider two model scenarios:  (i) a model that contains no orientational dependence for POPC-POPC interactions, and (ii) a model that includes explicit orientation dependence for POPC-POPC interactions. 

If we include  no orientation-dependent interactions between neighboring POPC molecules, the lipid-lipid Hamiltonian for a pure POPC bilayer is written
\begin{equation}
H_{popc-popc} = - \sum_{\left< i,j \right>} V_0^{popc}s_i s_j
\end{equation}
where the angled brackets indicate that the sum is taken over nearest neighbor pairs and
the coupling constant, $V_0^{popc}$, is a phenomenological parameter that is tuned
in such a way that the calculated phase transition temperature of pure POPC
matches the experimental value.  
To accomplish this, the average order parameter is calculated in a system of pure POPC for a range
of temperatures with a given value of $V_0^{popc}$.  
A plot of order parameter against temperature reveals a curve that
exhibits a sharp increase in order as temperature is decreased below a threshold temperature, $T_m$
which identifies $T_m$ as the phase transition
temperature.  $V_0^{popc}$ is phenomenologically  adjusted so that the simulated phase transition temperature
is in agreement with the experimentally-calculated value.
The numerical value for $V_0^{popc}$ for this case is shown in table~\ref{tpopc:}.

With two identical saturated chains, PSM is not expected to interact in an angular dependent
fashion among nearest neighbors.
Therefore, PSM:PSM interactions are written as:
\begin{equation} 
H_{psm-psm} = -\sum_{\left<i,j\right>} V_0^{psm} s_i s_j.
\label{hsmsm}
\end{equation}
$V_0^{psm}$ is the coupling constant tuned in pure
PSM mixtures to obtain a main chain phase transition temperature that
is identical to experiment.

\begin{figure}[htb]
\centering
   \includegraphics[width=85mm]{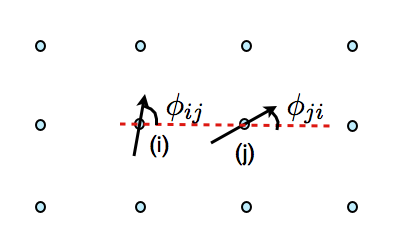}
      \caption{The angles $\phi_{ij}$ and $\phi_{ji}$ discussed in the text.  Black arrows represent
      the orientations of lipids at sites $i$ and $j$.  $\phi_{ij}$ and $\phi_{ji}$ are measured relative
      to a line connecting the lipid at site $i$ with its nearest neighbor at site $j$.}
      \label{theta_ij}
\end{figure}

For the case where we include an orientational degree of freedom for POPC molecules,  we alter the POPC chain-chain interaction Hamiltonian as follows.  We first recall that the concentration field variable $\psi_i (t) $ can range continuously from $-1$ representing pure POPC to
$+1$ representing pure PSM.
To incorporate the asymmetric nature of 
POPC-POPC interactions proposed above, we include an angular
dependence to this term in proportion to the amount of POPC present between interacting nearest
neighbors:
\begin{equation} 
\begin{split}
&H_{lipid-lipid} = -V_0 \sum_{\left<i,j\right>}s_is_j [1 + C(\psi_i, \psi_j)X(\phi_{ij}, \phi_{ji})] \\
& -\sum_{i=1}^{n_{lip}}V_1s_i\psi_i -
   \sum_{\langle
    ij\rangle}V_2\psi_i\psi_j.
\label{hpopcpopc}
\end{split}
\end{equation}
Here, $V_0$ is a function of fraction of POPC, $c_{popc}$, and the fraction of 
PSM, $c_{psm}$, present in the entire system:
\begin{equation}
V_0 = V_0^{popc}c_{popc} + V_0^{sm}c_{sm}.
\end{equation}
As the fraction of POPC at site $i$ is 
$ \frac{1}{2} (1 - \psi_i)$,
the local normalized fraction of POPC at neighboring sites $i$ and $j$ is
\begin{equation}
C(\psi_i, \psi_j) =  \frac{1}{4} (2-\psi_i -\psi_j).
\label{Cij}
\end{equation}
We model the angular contribution between nearest neighbor POPC pairs $i,j$ by the function
$X (\phi_{ij}, \phi_{ji}) $ of the relative POPC orientational angles $\phi_{ij}$ and $\phi_{ji}$ for molecules 
at sites $i$ and $j$
\begin{equation}
X(\phi_{ij}, \phi_{ji}) = \frac{\alpha_0}{2} ( \cos \phi_{ij} - \cos \phi_{ji} ).
\label{eq_x}
\end{equation}
The angles $\phi_{ij}$ and $\phi_{ji}$ are defined in Figure~\ref{theta_ij} 
and discussed in more detail in the next paragraph.
The second term in Eq \ref{hpopcpopc} couples the order parameter field to the
concentration field and the third term couples concentrations at
neighboring sites.  As described in~\cite{jcp}, $V_1$ and $V_2$ are estimated
directly from MD simulations.  The concentration dependence in the first term of 
Eq \ref{hpopcpopc} (also see Eq \ref{Cij})
ensures that interactions have an angular dependence only in proportion to the amount of
POPC at each site.  

In order to model the condition that a 
lipid interacts more favorably with a nearest neighbor
 if its saturated chain points towards that neighbor and less favorably if its unsaturated chain points towards
 that neighbor, the orientation function in Eq \ref{eq_x}
 is used.
 For this function, the orientations of POPC lipids at site $i$ and $j$ are illustrated with black arrows in
 Fig \ref{theta_ij}
which point from the top of the unsaturated chain towards the top of the saturated chain.
$\phi_{ij}$ and $\phi_{ji}$ represent the orientation of those vectors
 relative to a line between site $i$ to site $j$ (the red line in Fig \ref{theta_ij}).  The angle $\phi_{ij}$ is related to the orientation angle $\rho_i$, defined above and illustrated in Fig \ref{orientation}, by a constant that depends on the location of site $j$ relative to site $i$.  If the  two black arrows in Fig \ref{theta_ij}  point towards each other
 we interpret this as a saturated chain of a molecule 
  pointing towards the saturated chain of the neighboring molecule
  and $\phi_{ij}=0$ and $\phi_{ji}=\pi$.  This configuration is optimal and no energy penalty is imposed.
  However, if the two black arrows are pointed away from each other($\phi_i=\pi, \phi_j=0$),
we interpret this as two unsaturated chains pointing towards each other, and we  impose
the maximum energy penalty.
The constant $\alpha_0$ represents the strength of the angular dependence.   In this model, if $\alpha_0$ is greater than 1, the total energy interaction can include a 
repulsion between two, nearest neighbor chains. Since steric repulsions are implicitly present in this model through mean field statistical mechanical calculations and conformation sampling, it is not necessary or correct to include them in the Hamiltonian.

  \subsubsection{Lipid-CHOL Interaction Term,  $H_{lip-CHOL}$} 
As discussed above and described in previous publications~\cite{jcp,scmft1,scmft2}, lipid-CHOL
interactions have an angular dependence in the plane of the bilayer because the 
``smooth'' face has a tendency to induce more order in chains than the ``rough'' face
does.  The second term in the Hamiltonian accounts for this asymmetric interaction dependence:
\begin{equation}\label{lip-CHOL} H_{lip-CHOL} = -\sum_{i=1}^{n_{lip}}
  \sum_{j=1}^{n_{CHOL}}V_{lc}(1-\Delta
  \sin \theta_{i,j})s_i\end{equation} 
  The sums are over all lipid chains and nearest neighbor  CHOL molecules. 
 Each CHOL `rod' has a body coordinate system with a $y''$-axis along the length
of the rod starting at the center of mass, 
and an $x''$-axis extending from the center of mass in the direction
perpendicular to the rod on the `rough' side.
 $\theta_{i,j}$ is  defined as the angle between  the
  $y''$-axis of CHOL $i$ and
 vector connecting the center of mass of CHOL $i$ and the position of lipid $j$
 on the lattice\cite{scmft2}.   The coupling constant, $V_{lc}$, is found from from MD trajectories by linear regression analysis
  of CHOL-lipid chain interactions as a function of order parameter,
  as described in previous papers~\cite{scmft1,scmft2}.
  The parameter $\Delta$ was introduced in earlier work~\cite{scmft2} as a means to incorporate the fact
  that lipid chains are more energetically attracted to the the smooth side
  of CHOL~\cite{scmft2}.
Thus, it serves to represent the asymmetry of lipid-CHOL interactions
  in the x-y plane of the bilayer.
 Numerical values for $V_{lc}$ and $\Delta$ are 
 the same as used earlier in~\cite{scmft2}.  
  
  \subsubsection{CHOL-CHOL Interaction Term , $H_{CHOL-CHOL}$}

As in previous work,
CHOL molecules are cast as two-dimensional rods that diffuse through the order parameter
and concentration fields according to a
 Langevin equation
~\cite{scmft1,scmft2,jcp}.  As in earlier work we take
\begin{equation}\label{CHOL-CHOL} H_{CHOL-CHOL}
  =\sum_{i=1}^{n_{CHOL}}\sum_{j=1}^{n_{CHOL}}V_{cc}^r(r_{ij})V_{cc}^{\kappa}(\kappa_{ij})
\end{equation}
where sums are taken over all neighboring CHOL molecules~\cite{jcp,scmft1,scmft2}.
Here, $r_{ij}$ is the distance between two close CHOLs,  with indices $i$ and $j$.
$\kappa_{i}$ and $\kappa_j$ are angles made between a fixed direction on the lattice and a fixed
direction on the body coordinates of the $i^{th}$ and $j^{th}$ CHOLs and 
$\kappa_{ij} =\kappa_i - \kappa_j$ is the difference between the two.
From MD simulations and the success of previous models~\cite{jcp,scmft1,scmft2}, it can be 
surmised that a simple repulsive interaction is sufficient to model CHOL-CHOL
interactions:
\[V_{cc}^rV_{cc}^\kappa = \left\{ 
\begin{array}{l l}
  \epsilon \sin^2(\kappa_{ij})  & \quad \mbox r_{ij} \le L \\
  0 & \quad \mbox r_{ij}  > L \\ \end{array} \right. \]
where $\epsilon = 13 k_BT$ and $L $ is the rod length~\cite{scmft1,scmft2}.

\subsection{Mean Field Analysis}

After each time step, new configurations of CHOL positions and orientations and a new concentration field are generated.  Assuming that lipid chain order relaxes rapidly between timesteps, 
the new order parameter field is found in the mean-field approximation.  The 
underlying statistical mechanical partition function is:
\begin{equation}
Z_{tot} = \sum_{i=1}^{n_{lip}} \sum_{all confs} exp\left[-\frac{H_{tot}}{k_b T}\right]
\label{ztot}
\end{equation}
where $H_{tot}$ is given in Eq \ref{pham} and described in detail above.
Sums are over all possible configurations, represented by the order parameter
in Eq \ref{op}, over all lipids in the system.
It is not possible or practical to specify all single chain configurations, so we make use of a representative library
of configurations relevant to a bilayer environment that we obtain from MD simulations~\cite{scmft1,scmft2}.

For a given set of concentration field values, $\{{\psi_k}\}_{k=1}^{N}$,
the mean molecular field at site $i$ due to neighboring lipids, CHOL
molecules, and the concentration field has the form:
\begin{equation}
\begin{split}
&\Phi_i =\\
&-\sum_{j=1}^{\nu} < s_j>V_0[1 + C(\psi_i, \psi_j)X(\phi_{ij}, \phi_{ji})] \\
&+ c_i V_{lc} + V_1(C) \psi_i
\end{split}
\label{ppsi}
\end{equation}
where $\nu$ is the number of nearest neighbors to $i$ ($=4$ on a square lattice).
Once $\Phi_i$ is calculated at each site, it is used to solve the self-consistent
equations for the local order parameter values, within the mean-field approximation:
\begin{equation} \langle s_i \rangle = 
\frac{\sum_{all conf} s_c exp[\beta \Phi_i s_c]}{\sum_{all  conf} exp[\beta \Phi_i s_c]} 
\label{p-sc}
\end{equation}
and to find the mean-field free energy:
\begin{equation}
\begin{split}
&F = U - TS  \\
&=-\sum_{i=1}^{n_{lip}}  k_b T \ln Z_i 
+ H_{cc} - \frac{1}{2} \sum_{i=1}^{n_{lip}} \sum_{j=1}^{\nu} V_2 \psi_i \psi_j.
\end{split}
\end{equation}

\subsection{CHOL Diffusion}

The rotational and translational diffusion of CHOL molecules 
over the order parameter and concentration fields 
are modeled by Langevin equations:
\begin{equation} 
\label{lan1}
\frac{\partial \overrightarrow{r}_k}{\partial t} =
-M_r \frac{\partial F}{\partial\overrightarrow{r}_k}+\overrightarrow{\eta_k}\end{equation} and
 \begin{equation}
 \label{lan2}
 \frac{\partial {\omega}_k}{\partial t} =
 -M_{\omega} \frac{\partial F}{\partial{\omega}_k}+{\zeta_k}\end{equation} 
 where $\eta_k$ and $\zeta_k$ are stochastic noise 
 components and $M_r$ and $M_\omega$ are CHOL 
 mobilities.  $\overrightarrow{r}_k$ defines the x,y position 
 of the center of the $k^{th}$ CHOL and $\omega_k$ is the orientation of
 the CHOL body $x''-y''$ axes introduced above.  
 $\eta_k$ and $\zeta_k$ are thermal fluctuations modeled 
 as random variables that obey
 fluctuation-dissipation theorems.
 $M_r$ and $M_\omega$, the CHOL mobilities, 
 are related to MD-extracted diffusion constant, $D$, and the rotational 
 diffusion constant, $D_{rot}$, by $M_r=D/k_BT$ 
 and $M_\omega=D_{rot}/k_BT$~\cite{scmft1,scmft2}.
 Numerical values for constants are shown in Table~\ref{tpopc:}.

\subsection{\label{simulation}Simulation}

The system is initially given a random set of order parameter and concentration
field values at each lattice site and
a random distribution of CHOLs at the relevant concentration.
The system of self-consistent equations is solved
and the mean-field free energy is calculated.
Following this, the angle field, concentration field, and CHOL molecules are
updated by one time step.  The self-consistent
equations are solved again and the process is repeated for a total time of at least 1 microsecond.

\section{\label{app}Results}

In the application to POPC-PSM-CHOL mixtures, we consider a 100 by 100 square lattice
with each lattice site representing a whole molecule order parameter.
Molecular order parameters are averaged over both chains, so the field is constructed to represent
10,000 lipid molecules to which CHOL molecules are added.
POPC:PSM concentrations were simulated between
either 5\%to 95\% or 10\% to 90\%, depending on CHOL concentration,
in increments of 10\%.  Mixtures including CHOL were simulated
at CHOL concentrations ranging from 0-50\% in increments of 5\%.  
All simulated concentrations have been run for at least 1 microsecond at a temperature
of $303$ K.  To observe the effects of the addition of angular dependence, simulations
were ran with values of $\alpha_0$ equal to 0.0, 0.3, 0.5, 0.7, and 1.0.
Table~\ref{tpopc:} lists the numerical values for all
input parameters in the model. 
While our goal in this modeling work was to obtain as many of the model parameters from MD simulations as possible, in some cases the mapping process involved the use of simulation properties (e.g. molecular energies) that were quite noisy. Thus the model interaction parameters that we used are not necessarily unique, but they are representative of the set of model parameters that should be used in this type of coarse grained modeling. In addition to interaction parameters, the Mean Field model we use requires that the order parameters be symmetric about zero, whereas the MD library order parameter values are not symmetrically distributed due to the bilayer environment. As discussed in earlier work, we introduce order parameter
offsets~\cite{scmft1,scmft2}, $s_0^{popc}$, and $s_0^{psm}$ which are the mean values of the weighted chain order parameters found from MD simulations of pure POPC and PSM bilayers, respectively.
  These offsets are subtracted from each 
value of order parameter in the library, a necessary step to observe a temperature
dependent phase transition in the mean field approximation, in the absence of a symmetric library of chain configurations~\cite{scmft1,scmft2}.

\begin{table*}[ht]
\centering                          
\begin{tabular}{c c c c}            
\hline\hline                        
Parameter & Value & Method of Estimation & Comment \\ [0.5ex] 
\hline                              
$V_0^{popc}$ & 143, 160 KJ/mol & P &  calibrated for each $\alpha_0$ \\              
$V_0^{psm}$ & 63 KJ/mol& P & \\
$s_0^{popc}$  & .111 & MD& see~\cite{scmft1,scmft2} \\
$s_0^{psm}$ & .283 & MD & see~\cite{scmft1,scmft2} \\
$V_1$ & 20 KJ/mol & REG &  see \cite{jcp} \\
$V_2$ & 100 KJ/mol & REG  & see \cite{jcp}  \\ 
$V_{lc}$ & 3.0 KJ/mol & REG &see \cite{scmft1, scmft2} \\  
$\Delta$ & 2.0 & REG & anisotropy factor see \cite{scmft2} \\
$cc_{boundary}$ & 0.94 nm & MD & hardcore repulsive CHOL-CHOL cutoff  \\
$D_{lipid}$ & $10^{-12} m^2/s $ & APPROX & \\
$D_{r} $ & $10^{-12} m^2/s$ & APPROX & see \cite{scmft1, scmft2} \\
$D_{\omega} $ & $10*D_{r}$ & APPROX & see \cite{scmft1, scmft2} \\
$D_{\rho}$ & .07 $radians^2 / ns$ & REG & \\
$\alpha_0$ & 0.0, 1.0 & P & dimensionless \\ [1ex]         
\hline                              
\end{tabular}
\caption{Simulation Parameters.  P = Phenomenological. 
MD = calculated directly from MD simulations.
REG = linear regression approximation from MD simulations.
APPROX=Order of magnitude approximation from MD simulations.}
\label{tpopc:} 
\end{table*}

For the MD-based parameters in Table~\ref{tpopc:} and chain configuration libraries we used the following
simulation protocols:
The GROMACS simulation package
~\cite{berendsen,lindahl} was used to simulate a 200-lipid bilayer with compositions
37.5\% POPC, 37.5\%PSM and 25\% CHOL, solvated in 7211 SPC-E
water molecules.  The LINCS algorithm~\cite{hess} was used for bond constraints
with an integration time step of 2 fs.  Periodic boundary conditions were used in all
three dimensions.  Long range electrostatics were calculated using the PME
algorithm~\cite{essmann} with a real space cutoff of 10 \AA{}.  Van der Waals
interactions were cut off at 16 \AA{}.  The NPT ensemble was employed, allowing
the volume to fluctuate and the Parinello-Rahman pressure couling 
scheme~\cite{nose,parrinello}
was used with a constant pressure of 1 atm.  Systems were simulated for 
100 ns at a temperature of 313 K.

\begin{figure*} 
    \centering
      \subfigure[snapshot of concentration field]{\label{popc_psi_plot}
      \includegraphics[width=50mm]{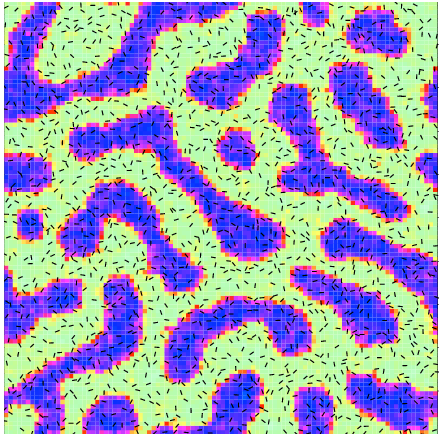}}
      \subfigure[snapshot of order parameter field]{\label{popc_op_plot}
      \includegraphics[width=50mm]{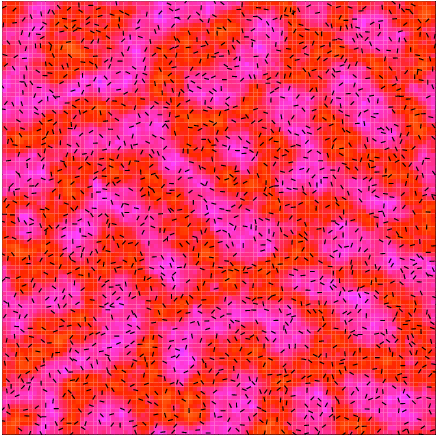}}
      \subfigure[Concentration Field scale] {\label{ppsiscale}
     \includegraphics[width=70mm]{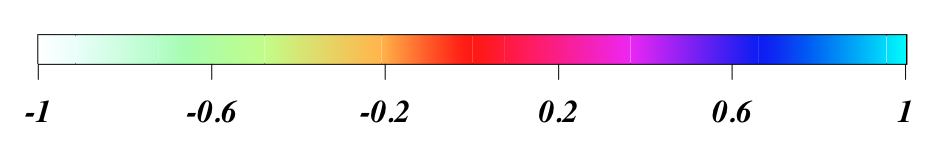}}
      \subfigure[order parameter scale] {\label{popscale}
     \includegraphics[width=70mm]{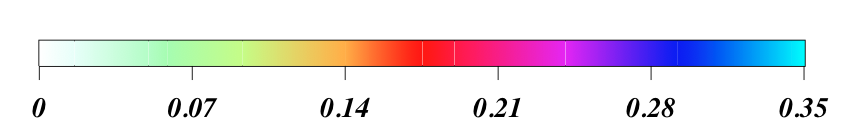}}
      \subfigure[Concentration field distribution]{\label{popc-psidist}
      \includegraphics[width=50mm]{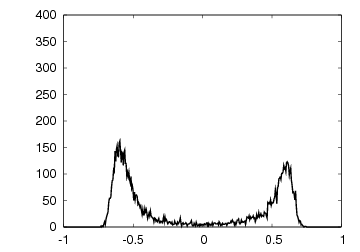}}
      \subfigure[Order parameter distribution]{\label{popc-opdist}
      \includegraphics[width=50mm]{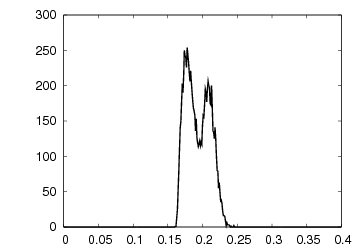}}
      \caption{Concentration field snapshot~\ref{popc_psi_plot} and scale~\ref{ppsiscale}, 
      Order parameter snapshot~\ref{popc_op_plot} and scale~\ref{popscale} (short black
      lines represent two-dimensional CHOL rods),
      concentration field distribution~\ref{popc-psidist} and
      order parameter distribution~\ref{popc-opdist} for the POPC:PSM:CHOL concentration 45:40:15.}
   \label{popc_dist}
\end{figure*}

The simulated results can be qualitatively visualized through two dimensional color density plots of concentration and order parameter fields. For quantitative analysis of the lateral organization we calculate binned distributions of order and concentration averaged over many snapshots.   Examples are shown in Fig \ref{popc_dist}.
The concentration field plot for the POPC:PSM:CHOL
concentration triplet 45:40:15 after 1 microsecond 
with $\alpha_0 = 0.0$ (no angular dependence), Fig \ref{popc_psi_plot},
exhibits a bimodal pattern of concentration with field value represented
by a color scheme shown in the key in~\ref{ppsiscale}.
The order parameter field snapshot for the same concentration after 1 microsecond is shown
in Fig \ref{popc_op_plot}.  This field exhibits a pattern that is similar to the pattern of the concentration field, with higher order parameter values associated with
higher concentrations of PSM and lower order 
parameter values associated with higher concentrations of POPC.  Figs \ref{popc-psidist} - \ref{popc-opdist} are plots of the distribution of values of the concentration and order fields over the lattice, respectively. This distribution, and others we describe here,  was made by placing the 10,000 lattice composition field values into
bins ranging in value from from -1 to +1(concentration)  or 0 to 0.5 (order) with width 0.01, for a single snapshot
after 1$\mu$s of simulation. However, we note that, based on both the color code for Fig \ref{popc_op_plot} and the narrowness of the distribution in Fig \ref{popc-opdist}, the variation of order over the lattice at this concentration is reduced compared to the variation in concentration over the lattice.

 \begin{figure*} 
    \centering
      \subfigure[$\alpha_0=0$]{\label{a0_0-50}
      \includegraphics[width=50mm]{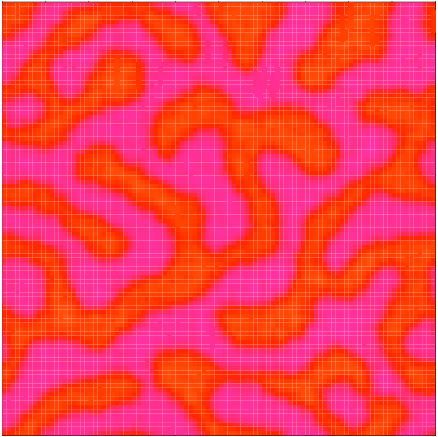}}
      \subfigure[$\alpha_0=1.0$]{\label{a10_0-50} 
      \includegraphics[width=50mm]{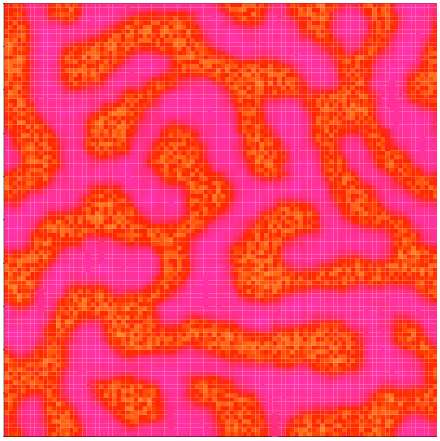}}
      \subfigure[$\alpha_0=0$]{\label{a0_20-40}
      \includegraphics[width=50mm]{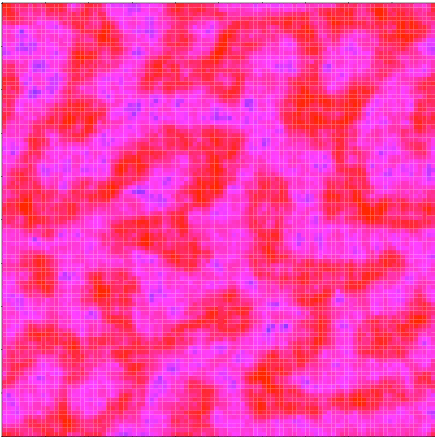}}
      \subfigure[$\alpha_0=1.0$]{\label{a10_20-40} 
      \includegraphics[width=50mm]{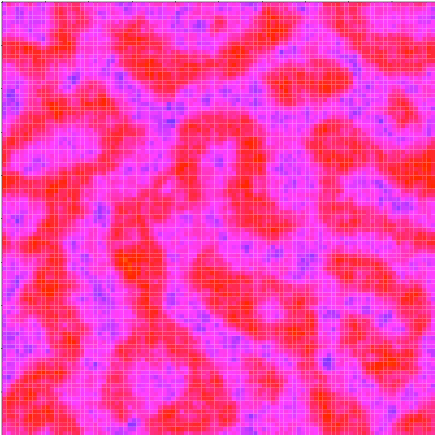}}
      \subfigure[order parameter scale] {\label{popscale2}
     \includegraphics[width=70mm]{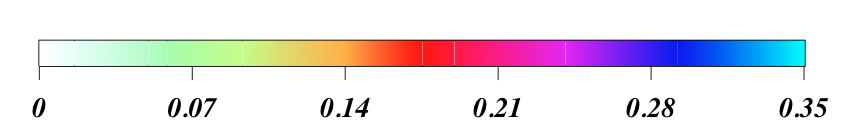}}
      \caption{Order parameter field snapshots for POPC:SM:CHOL concentrations
      50:50:0 (top) and 40:40:20 (bottom) for values of $\alpha_0=0.0$ and $\alpha_0=1.0$.
      CHOL molecules removed for clarity in bottom figures.  Scale is shown in~\ref{popscale2}}
   \label{alpha_difference}
\end{figure*}
 
To study the effect of POPC-POPC interaction angular dependence in the model,
we have run simulations with $\alpha_0$ values between 0.0 and 1.0.
The effect of increasing $\alpha_0$ is to very slightly increase the magnitude of lateral 
organizational patterns in the order parameter field at low CHOL concentration.  
Fig \ref{alpha_difference} illustrates the effect of $\alpha_0$ for two 
mixtures, POPC:PSM:CHOL 50:50:0 (Figs \ref{a10_0-50} and~\ref{a10_0-50})
and 40:40:20 (Figs \ref{a0_20-40} and~\ref{a10_20-40}). 
In both cases,
$\alpha_0 = 1.0$ leads to a slightly higher degree of lateral segregation in the order parameter field.
 Generally, we find that the incorporation of angular dependence leads to minor changes in the properties of the mixtures for CHOL concentrations less that about 20\%.  At CHOL concentrations closer to
 50\%, snapshots with $\alpha_0=0.0$ and
 $\alpha_0=1.0$ are indistinguishable. 
\begin{figure*}[htb]
 \centering
    \includegraphics [width=170mm]{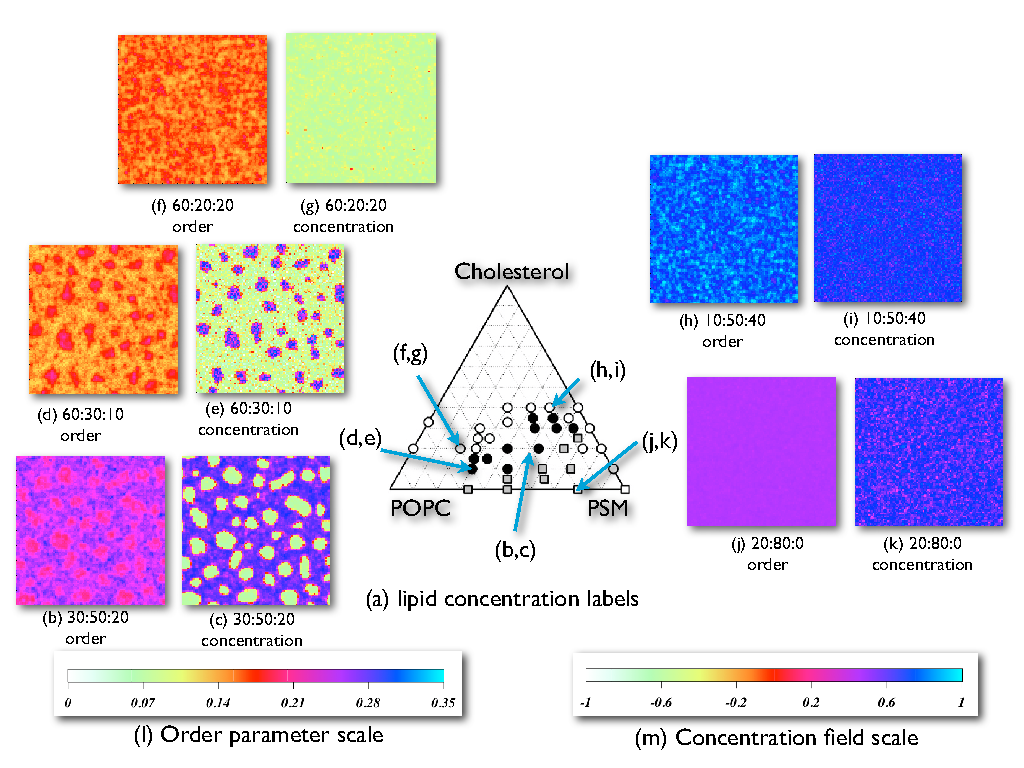}
    \caption{Experimentally derived phase plot \cite{veatch}, superimposed with labels
  marking the positions of the POPC:
  PSM:CHOL concentration fields and order parameter fields discussed in the text.
  In the simulation snapshots, CHOL molecules are removed for clarity.
  In all cases, $\alpha_0=1.0$.}
 \label{plabels}
 \end{figure*}
 
 By examining order parameter and concentration field distributions, and snapshots, we
 can locate points on a ternary triangle diagram where the model predicts bimodal distributions of order and concentration, which we then relate to experimental ternary bilayer triangle diagram regions.
 Fig \ref{plabels}(a) is a ternary plot published by Veatch and Keller~\cite{veatch}, 
 based on fluorescence microscopy for the temperature of 23$^\circ C$.
  Superimposed over their plot are arrows
linking simulated concentration points on the diagram to system snapshots in figures~\ref{plabels}(b)-~\ref{plabels}(k),
 for simulations carried out with $\alpha_0=1$.  In comparing our model to experimental data based on fluorescence microscopy, it is important to note that the simulation time scale, on the order of 1 $\mu s$,
 is much smaller than experimental time scales. Lateral inhomogeneities in this model as well as those in  
 other models based
 on the Cahn-Hilliard~\cite{chaiken, peng} equation, tend to increase in size over time under favorable conditions.   

Fig \ref{plabels}(b) shows the final order parameter snapshot for the POPC:PSM:CHOL
concentration 30:50:20 and Fig \ref{plabels}(c) shows the final concentration field snapshot
the same concentration.  This concentration corresponds to the point labeled (b,c) with an arrow in Fig \ref{plabels}(a).
We can see in the concentration field rounded regions
of light green color, indicating they are enriched with POPC.  They are embedded in a background
 of a dark blue color, indicating enrichment in PSM. The order parameter field color plot for this mixture, Fig \ref{plabels}(b), has a pattern that is similar to the concentration field plot, but note that the difference in the magnitude of the order parameters, the color scale difference, is reduced.

On the triangle diagram, experimental data points surrounding the point (b,c) are black circles,
which denote coexisting high and low order domains.  In the ternary plot of the same POPC-PSM-Chol system published by de Almeida \emph{et al}.~\cite{dealmeida},
this concentration point is considered to be within a three-phase region of coexisting gel, liquid disordered and 
liquid ordered phases.   Zhao \emph{et al}.~\cite{zhao}, argue that in some cases micron-sized separated regions may be
light induced~\cite{zhao}, but they do not rule out their existence
on a smaller size scale.

In Fig \ref{plabels}(d) we show the final order parameter field snapshot and in
Fig \ref{plabels}(e) we show  the final concentration field snapshot 
for the 60:30:10 mixture.  The corresponding point on the triangle diagram is labeled (d,e) by an arrow.
Again, we see the formation of 
higher ordered regions in the order parameter field coinciding
with PSM-rich regions in the concentration field and regions of lower order coinciding 
with POPC-rich regions.  
 The experimental data points on the triangle diagram in Fig \ref{plabels}(a) 
shows black dots at regions surrounding this concentration point, representing
the coexistence of liquid ordered and liquid disordered phases.  

In
Figs \ref{plabels}(f) and~\ref{plabels}(g), we show the final order parameter and concentration
field snapshots at the concentration triplet 60:20:20, labeled (f,g) with an arrow on the triangle diagram.  
For this point, where the POPC concentration is high, there are no clear bimodal patterns in either field.
However, CHOL does have an effect on the order, raising the values
of chain order without forming large-scale lateral inhomogeneity.  
At similar POPC:PSM ratios (3:1), but in the absence of CHOL (not shown in this figure)
we find that the order parameter field is quite uniform with an average
value of $\sim$0.15, with a very narrow distribution.
However, Figs \ref{plabels}(f) and~\ref{plabels}(g) show that  20 \% CHOL increases the order
into the range $\sim$0.16-0.17 with a wider distribution.  
In the order parameter field color density plot, there are diffuse regions that 
are more heavily populated by higher order and diffuse regions that are more heavily 
populated by lower order, but sharp demarcation between regions is absent.  
The experimental data points near point (f,g) are white circles that
represent a single liquid phase region.  De Almeida \emph{et al}.~\cite{dealmeida} and Pokorny
\emph{et al}.~\cite{pokorny} find evidence for coexistence of liquid ordered and liquid
disordered phases at this point on their POPC-PSM-CHOL triangle diagram.

The concentration triplet 10:50:40  is shown in Fig \ref{plabels}(h) and~\ref{plabels}(i)
and labeled (h,i) on the triangle diagram.
At this concentration there are no significant inhomogeneities in either
the order parameter or the concentration fields.  In this case there is a greater presence of PSM
and the result is a uniformly higher order parameter field and a uniform concentration field.
The presence of CHOL at this point
has the effect of inducing order on all of the chains.
The experimental data points surrounding
the simulated concentration are white circles that represent the presence of a
single liquid.  Pokorny \emph{et al}.~\cite{pokorny} and de Almeida \emph{et al}.~\cite{dealmeida}
also identify this region as a single liquid disordered phase.

Figs \ref{plabels}(j) and \ref{plabels}(k) show order parameter and concentration 
field final snapshots for
80\% PSM and 20\% POPC in the absence of CHOL. The corresponding concentration is
labeled (j,k) on the experimental diagram.
There is no discernible separation of order
parameter and concentration field values between high and low values.  
Both order parameter and concentration field distributions for this ratio are quite uniform.  
At this same concentration, the experimental data point shown
in Fig \ref{plabels}(a) is a white square, indicative of a single highly ordered gel phase.  
In our simulations, along the 0\% CHOL concentration axis, as POPC
concentration increases from 10\% to 90\%, the order parameter field evolves smoothly (within the resolution the number of simulations) from a high value, 
to low value with increasing POPC. Bimodalities in order are seen as shoulders rather than separated peaks, and one can discern subtle variations in snapshots, as seen for example in Fig \ref{alpha_difference}.  In the concentration field, however, there is a region of coexistence between PSM concentrations of 0.30 and 0.75 at 0\% CHOL.
At 50:50:0, the domains in the model are no longer circular, but percolate across the simulation box
(see Fig \ref{alpha_difference}).
The triangle diagram published by de Almeida \emph{et al.}~\cite{dealmeida} exhibits bimodal behavior along the 0\% CHOL 
concentration line, with a gel phase at the high PSM end,  liquid disordered phase at the
high POPC line, and a broad coexistence region between.


\begin{figure}[htb]
\centering
   \includegraphics[width=85mm]{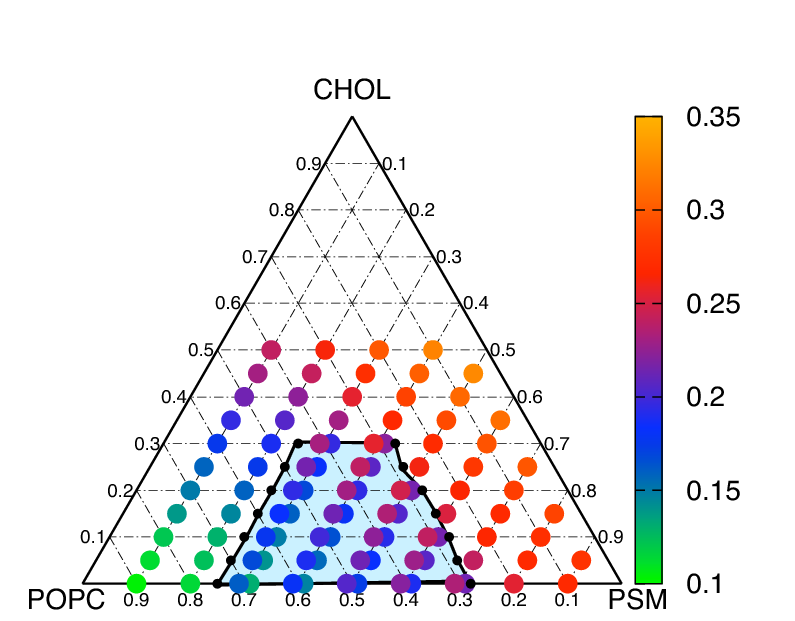}
      \caption{POPC-PSM-CHOL ternary phase diagram with $\alpha_0=1.0$.
      The shaded region
      indicates the presence of bimodality in the order parameter distributions.
      Dots indicate the average order parameter values, with color
      key to the right.  Two overlapping dots that appear
      in the bimodal region represent the average order parameter of each peak.}
      \label{a10popc_tern_plot}
\end{figure}

In Fig \ref{a10popc_tern_plot}, we collect all of the results of 1 
$\mu s$ simulations onto a triangle diagram based on order parameters for $\alpha_0=1.0$.
The triangle diagrams for other values of $\alpha_0$ are not qualitatively different.
The colored dots represent the average value of molecular order parameters, with the color of the dots denoting the magnitudes.
Within the shaded region, there are two dots at each simulated mixture point representing
the respective values at each peak in order parameter distributions which show bimodality.
The bimodal region itself is outlined  by a boundary contour that was approximated
by fitting distribution peaks to gaussians and interpolating where single peaks would
branch into double peaks.  The resolution of this boundary is limited by the number of 
simulations carried out. As expected, Fig \ref{a10popc_tern_plot} shows that increasing CHOL concentration has the effect of increasing the order 
parameter for the full system. 

\begin{figure*} 
    \centering
      \subfigure[5:80:15 order]{\label{p_op_15_5}\includegraphics[width=40mm]
      {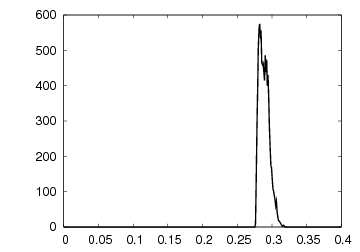}}
      \subfigure[5:80:15 concentration]{\label{p_psi_15_5}\includegraphics[width=40mm]
      {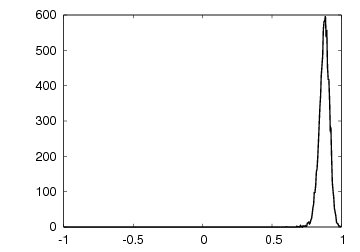}}
      \subfigure[15:70:15 order]{\label{p_op_15_15}\includegraphics[width=40mm]
      {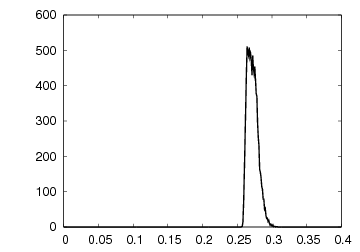}} 
      \subfigure[15:70:15 concentration]{\label{p_psi_15_15}\includegraphics[width=40mm]
      {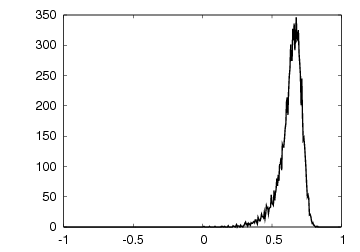}} 
      \subfigure[25:60:15 order]{\label{p_op_15_25}\includegraphics[width=40mm]
      {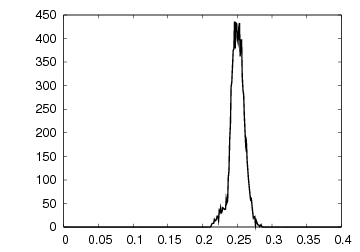}} 
      \subfigure[25:60:15 concentration]{\label{p_psi_15_25}\includegraphics[width=40mm]
      {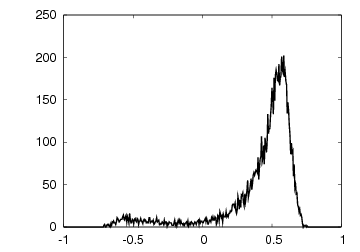}} 
            \subfigure[35:50:15 order]{\label{p_op_15_35}\includegraphics[width=40mm]
      {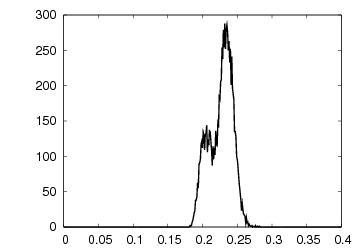}} 
      \subfigure[35:50:15 concentration]{\label{p_psi_15_35}\includegraphics[width=40mm]
      {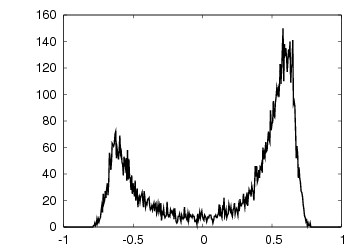}} 
            \subfigure[45:40:15 order]{\label{p_op_15_45}\includegraphics[width=40mm]
      {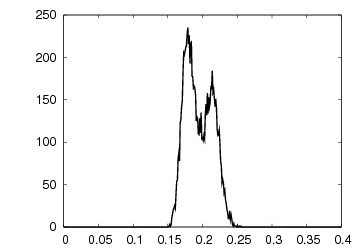}} 
      \subfigure[45:40:15 concentration]{\label{p_psi_15_45}\includegraphics[width=40mm]
      {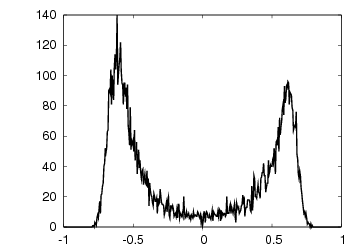}} 
            \subfigure[55:30:15 order]{\label{p_op_15_55}\includegraphics[width=40mm]
      {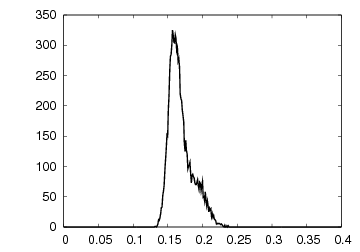}} 
      \subfigure[55:30:15 concentration]{\label{p_psi_15_55}\includegraphics[width=40mm]
      {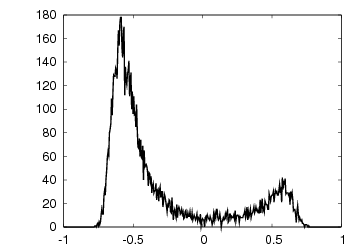}} 
            \subfigure[65:20:15 order]{\label{p_op_15_65}\includegraphics[width=40mm]
      {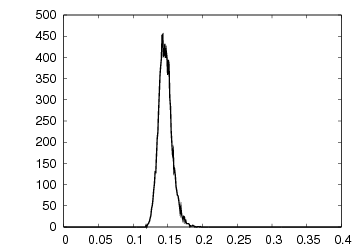}} 
      \subfigure[65:20:15 concentration]{\label{p_psi_15_65}\includegraphics[width=40mm]
      {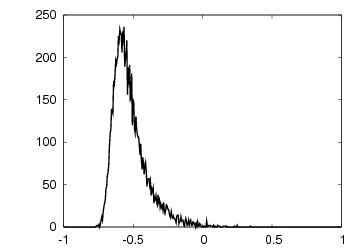}} 
            \subfigure[75:10:15 order]{\label{p_op_15_75}\includegraphics[width=40mm]
      {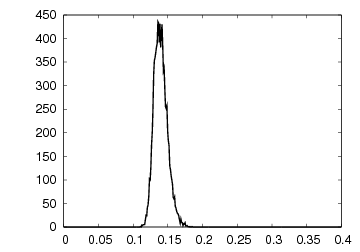}} 
      \subfigure[75:10:15 concentration]{\label{p_psi_15_75}\includegraphics[width=40mm]
      {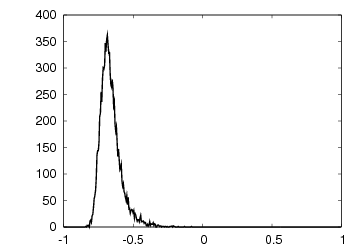}} 
      \caption{Distribution plots of
       order parameter
      and concentration field for the POPC:PSM:CHOL concentrations, 
      for $\alpha_0 = 1.0$, labeled for the last snapshot of 1$\mu s$
      simulations along the 15\% CHOL concentration line.}
   \label{snapshots_15_popc}
\end{figure*}

As can be seen in the ternary figure, the patterns of the colored dots exhibit a shift from high order parameter
value (orange, yellow) in PSM-rich regions to low order parameter (green, blue) in POPC-rich regions.  
At concentrations of CHOL below about 0.3,  the order parameter
and concentration fields separate into regions rich in POPC and regions rich in PSM, as was discussed above.
To quantitatively analyze this behavior, we consider the 15\% CHOL concentration line in the ternary
plot, whose distributions of order parameter and concentration fields are 
shown in Figs \ref{p_op_15_5}-\ref{p_psi_15_75}.
At low POPC concentration, 5:80:15 (Figs \ref{p_op_15_5} and \ref{p_psi_15_5}), 
the order fields have a single peak and the average
value occurs at approximately 0.28 (a  yellow dot
in Fig \ref{a10popc_tern_plot}).  As we move toward higher POPC concentrations, the order parameter
value is reduced.  At the concentration 35:60:15 (Figs \ref{p_op_15_15} and \ref{p_psi_15_15}), 
we start to see the emergence of
a second peak in concentration around 0.2.  Moving further into the bimodal region, this concentration field peak grows and
eventually dominates the distribution.  Above 55:30:15 (Figs \ref{p_op_15_55} and \ref{p_psi_15_55}), 
again, only a single peak is visible in the distribution. Note that in all cases the order  parameter field exhibits a much more subtle level of bimodality, through shoulders in a single peak rather than separated peaks.
The addition of angular dependence does not change the distributions significantly.
\begin{figure} 
    \centering
      \subfigure[10:40:50 order]{\label{p_op_50_10}\includegraphics[width=40mm]
      {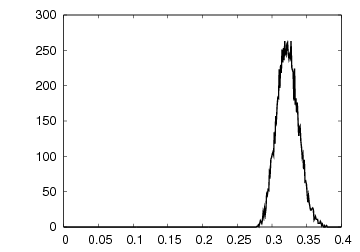}}
      \subfigure[10:40:50 concentration]{\label{p_psi_50_10}\includegraphics[width=40mm]
      {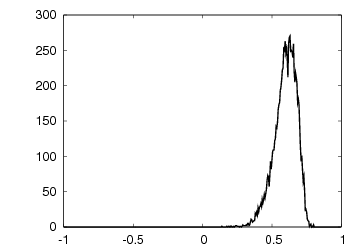}}
      \subfigure[20:30:50 order]{\label{p_op_50_20}\includegraphics[width=40mm]
      {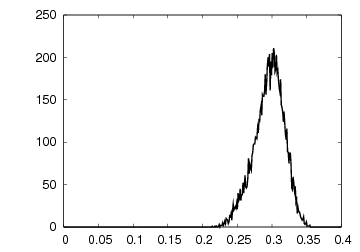}} 
      \subfigure[20:30:50 concentration]{\label{p_psi_50_20}\includegraphics[width=40mm]
      {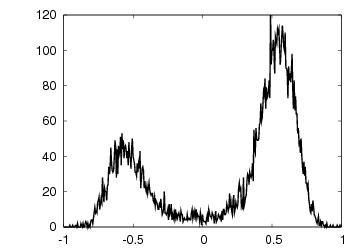}} 
      \subfigure[30:20:50 order]{\label{p_op_50_30}\includegraphics[width=40mm]
      {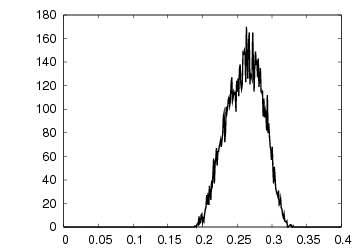}} 
      \subfigure[30:20:50 concentration]{\label{p_psi_50_30}\includegraphics[width=40mm]
      {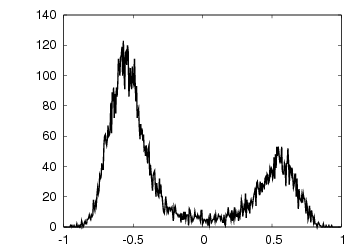}} 
            \subfigure[40:10:50 order]{\label{p_op_50_40}\includegraphics[width=40mm]
      {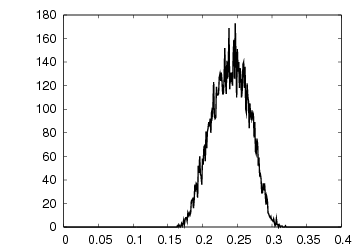}} 
      \subfigure[40:10:50 concentration]{\label{p_psi_50_40}\includegraphics[width=40mm]
      {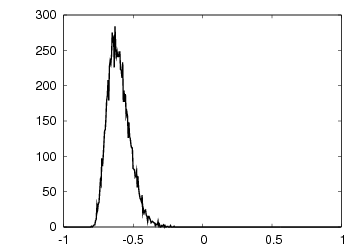}} 
                 \caption{Distribution plots of
       order parameter
      and concentration field
      for the POPC:PSM:CHOL concentrations, for $\alpha_0 = 1.0$,
       labeled for the last snapshot of 1$\mu s$
      simulations along the 50\% CHOL concentration line.}
   \label{snapshots_50_popc}
\end{figure}

As CHOL increases above 15\% we find that the order parameter bimodal peaks begin to merge, 
while the concentration field distribution remains bimodal.  The distributions of the order and concentration fields  at
50\% CHOL are shown in
Fig \ref{snapshots_50_popc}.  At high PSM concentrations, 10:40:50 (Figs \ref{p_op_50_10} and 
\ref{p_psi_50_10}) there is a single peak in both the concentration field and the order parameter field.
This is also true at high POPC concentrations, 40:10:50 (Figs \ref{p_op_50_40} and \ref{p_psi_50_40}).
However, between these two extremes, 
(Figs \ref{p_op_50_20} and \ref{p_psi_50_20} and 
Figs \ref{p_op_50_30} and \ref{p_psi_50_30}), the concentration
field distributions split into two clearly separate peaks, but {\em the order parameter
distributions still show just a single peak}. In this model, we find that the presence of CHOL at moderate to high concentration (above
30\% CHOL concentration) has the effect of ordering lipid chains of both PSM and POPC 
to such a degree that we are unable to distinguish
between POPC and PSM by chain order.

\section{\label{discussion}Discussion}

We have presented a quantitative model for ternary mixtures of POPC, PSM, and CHOL. The model represents a high level of coarse graining that allows us to simulate many different mixtures on relatively large lattices for relatively long times. The results presented in the form of distributions of order and concentrations, and in the form of final snapshots, fit well with existing experimental data. The degree of phenomenology is necessarily greater than our earlier work~\cite{jcp} but every effort was made to use input from MD simulations where possible.  As we found in earlier work, the order parameter field and concentration field distributions vary according to the mean field approximation, and are expected to provide an accurate description of the system as long as one is not near a critical point. In the model, 
as relative concentrations of POPC, PSM, and CHOL are changed, bimodalities in the concentration distributions can be found over a portion of the triangle diagram as illustrated in  Fig \ref{a10popc_tern_plot}.  
For intermediate CHOL concentrations, bimodalities in order parameter field distributions 
tend to parallel those found in concentration field distributions.  However, in the POPC-PSM-CHOL results, differences in chain order between adjacent domains are generally smaller in magnitude  when compared to results from the same model applied to DOPC-SSM-CHOL \cite{jcp}.
At higher CHOL concentrations (above about 30\%) order
parameter bimodality  is not found even for mixtures that exhibit concentration field bimodality.

\begin{figure*} 
    \centering
      \subfigure[DOPC 60:30:10]{\label{dopc-10-60}\includegraphics[width=50mm]
      {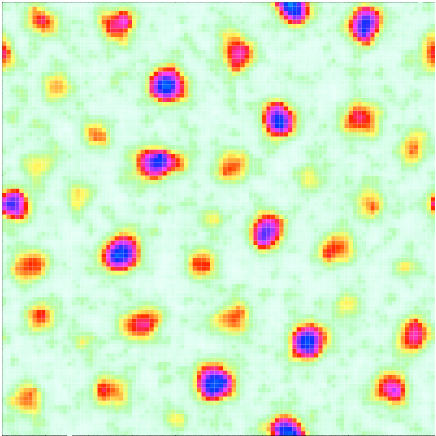}}
      \subfigure[POPC 60:30:10 $\alpha_0=0.0$]{\label{popc-10-60-0}\includegraphics[width=50mm]
      {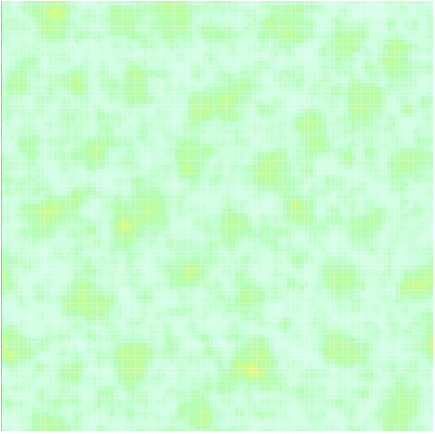}}
      \subfigure[POPC 60:30:10 $\alpha_0=1.0$]{\label{popc-10-60-10}\includegraphics[width=50mm]
      {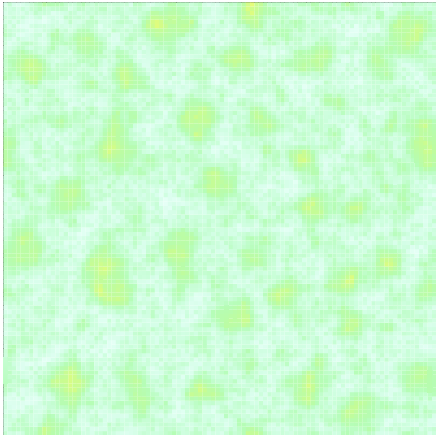}} 
            \subfigure[Scale]{\label{compare_scale}\includegraphics[width=70mm]
      {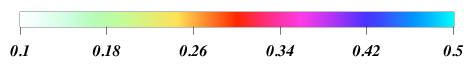}} 
                 \caption{Order parameter field final snapshots for DOPC:SM:CHOL
                 60:30:10 and POPC:PSM:CHOL mixtures at the same concentration ratios
                 with $\alpha_0=0.0$ and $\alpha_0=1.0$.  Scale shown in \ref{compare_scale}.}
   \label{comparison}
\end{figure*}

A key difference between  the results of this paper and our previous implementation of the SCMFT model for DOPC-SSM-CHOL mixtures \cite{jcp} is the reduced area occupied by the bimodal regions the POPC-PSM-CHOL
ternary plot (see \cite{jcp}),compared to the bimodal regions in the DOPC-SSM-CHOL ternary
plots.
In the DOPC system, the bimodal region is wider, 
implying that the DOPC system can more readily separate in to high/low
order regions and the inclusion of a saturated chain in POPC reduces the size of the regions over which separated domains can form.
 Fig \ref{comparison} illustrates this comparative difference
60:30:10 mixtures (Fig \ref{dopc-10-60}).
In the DOPC mixture, small rounded domains have formed that are rich in
ordered SSM and are surrounded by a DOPC background with a lower order parameter value.
In the POPC mixtures, the color density plot shows that the difference in the order parameters between the regions is practically zero. 
Therefore, a prediction of this model is that molecular chain order differences may be quite subtle and even undetectable whereas concentration field distributions still exhibit bimodality, or phase separation.

We note that, while the inclusion of an orientational interaction for asymmetrical POPC molecules is a reasonable addition to the model, the results in our work are only weakly dependent on this interaction. The interaction function that included POPC orientation, Eq.~\ref{hpopcpopc}, was chosen as a simple way to express this interaction. It is possible that a different choice for the orientational energy expression could produce different results when the concentration of POPC is large.

In summary, our extended model for ternary component lipid
bilayers of POPC-PSM- CHOL shows good agreement with experiment, and provides possible new insights into the atomic level compositional and order distributions.
This SCMFT model can be utilized to describe any system consisting of two long-chain
lipids in a ternary mixture with CHOL, independent of chain type, length, or level
of unsaturation, while also taking into account any angular dependencies.

\begin{acknowledgments}
SP thanks Gerald Feigenson for helpful discussions.
PT, GZ, and HLS acknowledge support from NIH Grant Number PHS 2 PN2 EY016570B from the 
through the NIH Roadmap for Medical Research.  Author SAP acknowledge support from NIH 
grant 1R01GM086707-01A1.
\end{acknowledgments}

\break


\begin{thebibliography}{75}
\expandafter\ifx\csname natexlab\endcsname\relax\def\natexlab#1{#1}\fi
\expandafter\ifx\csname bibnamefont\endcsname\relax
  \def\bibnamefont#1{#1}\fi
\expandafter\ifx\csname bibfnamefont\endcsname\relax
  \def\bibfnamefont#1{#1}\fi
\expandafter\ifx\csname citenamefont\endcsname\relax
  \def\citenamefont#1{#1}\fi
\expandafter\ifx\csname url\endcsname\relax
  \def\url#1{\texttt{#1}}\fi
\expandafter\ifx\csname urlprefix\endcsname\relax\def\urlprefix{URL }\fi
\providecommand{\bibinfo}[2]{#2}
\providecommand{\eprint}[2][]{\url{#2}}

\bibitem[{\citenamefont{Veatch and Keller}(2005)}]{veatch}
\bibinfo{author}{\bibfnamefont{S.~L.} \bibnamefont{Veatch}} \bibnamefont{and}
  \bibinfo{author}{\bibfnamefont{S.~L.} \bibnamefont{Keller}},
  \bibinfo{journal}{Phys. Rev. Lett.} \textbf{\bibinfo{volume}{94}},
  \bibinfo{pages}{148101} (\bibinfo{year}{2005}).

\bibitem[{\citenamefont{Haluska et~al.}(2008)\citenamefont{Haluska,
  Schr\"{o}der, Didier, Heissler, Duportail, M\'{e}ly, and Marques}}]{haluska}
\bibinfo{author}{\bibfnamefont{C.~K.} \bibnamefont{Haluska}},
  \bibinfo{author}{\bibfnamefont{A.~P.} \bibnamefont{Schr\"{o}der}},
  \bibinfo{author}{\bibfnamefont{P.}~\bibnamefont{Didier}},
  \bibinfo{author}{\bibfnamefont{D.}~\bibnamefont{Heissler}},
  \bibinfo{author}{\bibfnamefont{G.}~\bibnamefont{Duportail}},
  \bibinfo{author}{\bibfnamefont{Y.}~\bibnamefont{M\'{e}ly}}, \bibnamefont{and}
  \bibinfo{author}{\bibfnamefont{C.}~\bibnamefont{Marques}},
  \bibinfo{journal}{Biophys. J.} \textbf{\bibinfo{volume}{95}}
  (\bibinfo{year}{2008}).

\bibitem[{\citenamefont{Kahya et~al.}(2003)\citenamefont{Kahya, Scherfeld,
  Bacia, Poolman, and Schwille}}]{kahya}
\bibinfo{author}{\bibfnamefont{N.}~\bibnamefont{Kahya}},
  \bibinfo{author}{\bibfnamefont{D.}~\bibnamefont{Scherfeld}},
  \bibinfo{author}{\bibfnamefont{K.}~\bibnamefont{Bacia}},
  \bibinfo{author}{\bibfnamefont{B.}~\bibnamefont{Poolman}}, \bibnamefont{and}
  \bibinfo{author}{\bibfnamefont{P.}~\bibnamefont{Schwille}},
  \bibinfo{journal}{J. Biol. Chem.} \textbf{\bibinfo{volume}{278}},
  \bibinfo{pages}{28109} (\bibinfo{year}{2003}).

\bibitem[{\citenamefont{Veatch and Keller}(2002)}]{veatch3}
\bibinfo{author}{\bibfnamefont{S.~L.} \bibnamefont{Veatch}} \bibnamefont{and}
  \bibinfo{author}{\bibfnamefont{S.~L.} \bibnamefont{Keller}},
  \bibinfo{journal}{Phys. Rev. Lett.} \textbf{\bibinfo{volume}{89}},
  \bibinfo{pages}{268101} (\bibinfo{year}{2002}).

\bibitem[{\citenamefont{Stottrup et~al.}(2005)\citenamefont{Stottrup, Stevens,
  and Keller}}]{stottrup}
\bibinfo{author}{\bibfnamefont{B.}~\bibnamefont{Stottrup}},
  \bibinfo{author}{\bibfnamefont{D.}~\bibnamefont{Stevens}}, \bibnamefont{and}
  \bibinfo{author}{\bibfnamefont{S.}~\bibnamefont{Keller}},
  \bibinfo{journal}{Biophys. J.} \textbf{\bibinfo{volume}{88}},
  \bibinfo{pages}{269} (\bibinfo{year}{2005}).

\bibitem[{\citenamefont{Aittoniemi et~al.}(2007)\citenamefont{Aittoniemi,
  Niemel\"{a}, Hyv\"{o}nen, Karttunen, and Vattulainen}}]{aittoniemi}
\bibinfo{author}{\bibfnamefont{J.}~\bibnamefont{Aittoniemi}},
  \bibinfo{author}{\bibfnamefont{P.~S.} \bibnamefont{Niemel\"{a}}},
  \bibinfo{author}{\bibfnamefont{M.~T.} \bibnamefont{Hyv\"{o}nen}},
  \bibinfo{author}{\bibfnamefont{M.}~\bibnamefont{Karttunen}},
  \bibnamefont{and}
  \bibinfo{author}{\bibfnamefont{I.}~\bibnamefont{Vattulainen}},
  \bibinfo{journal}{Biophysical Journal} \textbf{\bibinfo{volume}{92}},
  \bibinfo{pages}{1125} (\bibinfo{year}{2007}).

\bibitem[{\citenamefont{Zhang et~al.}(2007)\citenamefont{Zhang, Bhide, and
  Berkowitz}}]{zhang}
\bibinfo{author}{\bibfnamefont{Z.}~\bibnamefont{Zhang}},
  \bibinfo{author}{\bibfnamefont{S.~Y.} \bibnamefont{Bhide}}, \bibnamefont{and}
  \bibinfo{author}{\bibfnamefont{M.~L.} \bibnamefont{Berkowitz}},
  \bibinfo{journal}{The Journal of Physical Chemistry}
  \textbf{\bibinfo{volume}{111}}, \bibinfo{pages}{12888}
  (\bibinfo{year}{2007}).

\bibitem[{\citenamefont{de~Almeida et~al.}(2007)\citenamefont{de~Almeida,
  Borst, Fedorov, Prieto, and Visser}}]{dealmeida2}
\bibinfo{author}{\bibfnamefont{R.~F.~M.} \bibnamefont{de~Almeida}},
  \bibinfo{author}{\bibfnamefont{J.}~\bibnamefont{Borst}},
  \bibinfo{author}{\bibfnamefont{A.}~\bibnamefont{Fedorov}},
  \bibinfo{author}{\bibfnamefont{M.}~\bibnamefont{Prieto}}, \bibnamefont{and}
  \bibinfo{author}{\bibfnamefont{A.~J. W.~G.} \bibnamefont{Visser}},
  \bibinfo{journal}{Biophysical Journal} \textbf{\bibinfo{volume}{93}},
  \bibinfo{pages}{539} (\bibinfo{year}{2007}).

\bibitem[{\citenamefont{Lindblom et~al.}(2006)\citenamefont{Lindblom,
  Or\"{a}dd, and Filippov}}]{lindblom}
\bibinfo{author}{\bibfnamefont{G.}~\bibnamefont{Lindblom}},
  \bibinfo{author}{\bibfnamefont{G.}~\bibnamefont{Or\"{a}dd}},
  \bibnamefont{and} \bibinfo{author}{\bibfnamefont{A.}~\bibnamefont{Filippov}},
  \bibinfo{journal}{Chemistry and Physics of Lipids}
  \textbf{\bibinfo{volume}{141}}, \bibinfo{pages}{179} (\bibinfo{year}{2006}).

\bibitem[{\citenamefont{Zhao et~al.}(2007{\natexlab{a}})\citenamefont{Zhao, Wu,
  Heberle, Mills, Klawitter, Huang, Costanza, and Feigenson}}]{zhao}
\bibinfo{author}{\bibfnamefont{J.}~\bibnamefont{Zhao}},
  \bibinfo{author}{\bibfnamefont{J.}~\bibnamefont{Wu}},
  \bibinfo{author}{\bibfnamefont{F.~A.} \bibnamefont{Heberle}},
  \bibinfo{author}{\bibfnamefont{T.~T.} \bibnamefont{Mills}},
  \bibinfo{author}{\bibfnamefont{P.}~\bibnamefont{Klawitter}},
  \bibinfo{author}{\bibfnamefont{G.}~\bibnamefont{Huang}},
  \bibinfo{author}{\bibfnamefont{G.}~\bibnamefont{Costanza}}, \bibnamefont{and}
  \bibinfo{author}{\bibfnamefont{G.~W.} \bibnamefont{Feigenson}},
  \bibinfo{journal}{Biochimica et Biophysica Acta}
  \textbf{\bibinfo{volume}{1768}}, \bibinfo{pages}{2764}
  (\bibinfo{year}{2007}{\natexlab{a}}).

\bibitem[{\citenamefont{Collins and Keller}(2008)}]{collins}
\bibinfo{author}{\bibfnamefont{M.~D.} \bibnamefont{Collins}} \bibnamefont{and}
  \bibinfo{author}{\bibfnamefont{S.~L.} \bibnamefont{Keller}},
  \bibinfo{journal}{PNAS} \textbf{\bibinfo{volume}{105}}, \bibinfo{pages}{124}
  (\bibinfo{year}{2008}).

\bibitem[{\citenamefont{Heberle et~al.}(2005)\citenamefont{Heberle, Buboltz,
  Stringer, and Feigenson}}]{heberle}
\bibinfo{author}{\bibfnamefont{F.~A.} \bibnamefont{Heberle}},
  \bibinfo{author}{\bibfnamefont{J.~T.} \bibnamefont{Buboltz}},
  \bibinfo{author}{\bibfnamefont{D.}~\bibnamefont{Stringer}}, \bibnamefont{and}
  \bibinfo{author}{\bibfnamefont{G.~W.} \bibnamefont{Feigenson}},
  \bibinfo{journal}{Biochimica et Biophysica Acta}
  \textbf{\bibinfo{volume}{1746}}, \bibinfo{pages}{186} (\bibinfo{year}{2005}).

\bibitem[{\citenamefont{Moore et~al.}(2006)\citenamefont{Moore, Snyder, Rerek,
  and Mendelsohn}}]{moore}
\bibinfo{author}{\bibfnamefont{D.~J.} \bibnamefont{Moore}},
  \bibinfo{author}{\bibfnamefont{R.~G.} \bibnamefont{Snyder}},
  \bibinfo{author}{\bibfnamefont{M.~E.} \bibnamefont{Rerek}}, \bibnamefont{and}
  \bibinfo{author}{\bibfnamefont{R.}~\bibnamefont{Mendelsohn}},
  \bibinfo{journal}{J. Phys. Chem.} \textbf{\bibinfo{volume}{110}},
  \bibinfo{pages}{2378} (\bibinfo{year}{2006}).

\bibitem[{\citenamefont{Zheng et~al.}(2007)\citenamefont{Zheng, McQuaw, Ewing,
  and Winograd}}]{zheng}
\bibinfo{author}{\bibfnamefont{L.}~\bibnamefont{Zheng}},
  \bibinfo{author}{\bibfnamefont{C.~M.} \bibnamefont{McQuaw}},
  \bibinfo{author}{\bibfnamefont{A.~G.} \bibnamefont{Ewing}}, \bibnamefont{and}
  \bibinfo{author}{\bibfnamefont{N.}~\bibnamefont{Winograd}},
  \bibinfo{journal}{J. Am. Chem. Soc.} \textbf{\bibinfo{volume}{129}},
  \bibinfo{pages}{15730} (\bibinfo{year}{2007}).

\bibitem[{\citenamefont{Zhao et~al.}(2007{\natexlab{b}})\citenamefont{Zhao, Wu,
  Shao, Kong, Jain, Hunt, and Feigenson}}]{zhao2}
\bibinfo{author}{\bibfnamefont{J.}~\bibnamefont{Zhao}},
  \bibinfo{author}{\bibfnamefont{J.}~\bibnamefont{Wu}},
  \bibinfo{author}{\bibfnamefont{H.}~\bibnamefont{Shao}},
  \bibinfo{author}{\bibfnamefont{F.}~\bibnamefont{Kong}},
  \bibinfo{author}{\bibfnamefont{N.}~\bibnamefont{Jain}},
  \bibinfo{author}{\bibfnamefont{G.}~\bibnamefont{Hunt}}, \bibnamefont{and}
  \bibinfo{author}{\bibfnamefont{G.}~\bibnamefont{Feigenson}},
  \bibinfo{journal}{Biochimica et Biophysica Acta}
  \textbf{\bibinfo{volume}{1768}}, \bibinfo{pages}{2777}
  (\bibinfo{year}{2007}{\natexlab{b}}).

\bibitem[{\citenamefont{Filippov et~al.}(2007)\citenamefont{Filippov,
  Or\"{a}dd, and Lindblom}}]{filippov1}
\bibinfo{author}{\bibfnamefont{A.}~\bibnamefont{Filippov}},
  \bibinfo{author}{\bibfnamefont{G.}~\bibnamefont{Or\"{a}dd}},
  \bibnamefont{and} \bibinfo{author}{\bibfnamefont{G.}~\bibnamefont{Lindblom}},
  \bibinfo{journal}{Biophysical Journal} \textbf{\bibinfo{volume}{93}},
  \bibinfo{pages}{3182} (\bibinfo{year}{2007}).

\bibitem[{\citenamefont{Filippov et~al.}(2003)\citenamefont{Filippov,
  Or\"{a}dd, and Lindblom}}]{filippov2}
\bibinfo{author}{\bibfnamefont{A.}~\bibnamefont{Filippov}},
  \bibinfo{author}{\bibfnamefont{G.}~\bibnamefont{Or\"{a}dd}},
  \bibnamefont{and} \bibinfo{author}{\bibfnamefont{G.}~\bibnamefont{Lindblom}},
  \bibinfo{journal}{Biophysical Journal} \textbf{\bibinfo{volume}{84}},
  \bibinfo{pages}{3079} (\bibinfo{year}{2003}).

\bibitem[{\citenamefont{Cicuta et~al.}(2007)\citenamefont{Cicuta, Keller, and
  Veatch}}]{cicuta}
\bibinfo{author}{\bibfnamefont{P.}~\bibnamefont{Cicuta}},
  \bibinfo{author}{\bibfnamefont{S.~L.} \bibnamefont{Keller}},
  \bibnamefont{and} \bibinfo{author}{\bibfnamefont{S.~L.}
  \bibnamefont{Veatch}}, \bibinfo{journal}{The Journal of Physical Chemistry B}
  \textbf{\bibinfo{volume}{111}}, \bibinfo{pages}{3328} (\bibinfo{year}{2007}).

\bibitem[{\citenamefont{Castro et~al.}(2007)\citenamefont{Castro, de~Almeida,
  Silva, Fedorov, and Prieto}}]{castro}
\bibinfo{author}{\bibfnamefont{B.~M.} \bibnamefont{Castro}},
  \bibinfo{author}{\bibfnamefont{R.~F.~M.} \bibnamefont{de~Almeida}},
  \bibinfo{author}{\bibfnamefont{L.~C.} \bibnamefont{Silva}},
  \bibinfo{author}{\bibfnamefont{A.}~\bibnamefont{Fedorov}}, \bibnamefont{and}
  \bibinfo{author}{\bibfnamefont{M.}~\bibnamefont{Prieto}},
  \bibinfo{journal}{Biophysical Journal} \textbf{\bibinfo{volume}{93}},
  \bibinfo{pages}{1639} (\bibinfo{year}{2007}).

\bibitem[{\citenamefont{Bakht et~al.}(2007)\citenamefont{Bakht, Pathak, and
  London}}]{bakht}
\bibinfo{author}{\bibfnamefont{O.}~\bibnamefont{Bakht}},
  \bibinfo{author}{\bibfnamefont{P.}~\bibnamefont{Pathak}}, \bibnamefont{and}
  \bibinfo{author}{\bibfnamefont{E.}~\bibnamefont{London}},
  \bibinfo{journal}{Biophysical Journal} \textbf{\bibinfo{volume}{93}},
  \bibinfo{pages}{4307} (\bibinfo{year}{2007}).

\bibitem[{\citenamefont{Frazier et~al.}(2007)\citenamefont{Frazier, Wright,
  Pokorny, and Almeida}}]{frazier}
\bibinfo{author}{\bibfnamefont{M.~L.} \bibnamefont{Frazier}},
  \bibinfo{author}{\bibfnamefont{J.~R.} \bibnamefont{Wright}},
  \bibinfo{author}{\bibfnamefont{A.}~\bibnamefont{Pokorny}}, \bibnamefont{and}
  \bibinfo{author}{\bibfnamefont{P.~F.~F.} \bibnamefont{Almeida}},
  \bibinfo{journal}{Biophysical Journal} \textbf{\bibinfo{volume}{92}},
  \bibinfo{pages}{2422} (\bibinfo{year}{2007}).

\bibitem[{\citenamefont{Ursell et~al.}(2009)\citenamefont{Ursell, Klug, and
  Phillips}}]{ursell}
\bibinfo{author}{\bibfnamefont{T.~S.} \bibnamefont{Ursell}},
  \bibinfo{author}{\bibfnamefont{W.~S.} \bibnamefont{Klug}}, \bibnamefont{and}
  \bibinfo{author}{\bibfnamefont{R.}~\bibnamefont{Phillips}},
  \bibinfo{journal}{PNAS}  (\bibinfo{year}{2009}).

\bibitem[{\citenamefont{Veatch et~al.}(2007)\citenamefont{Veatch, Soubias,
  Keller, and Gawsrisch}}]{veatch4}
\bibinfo{author}{\bibfnamefont{S.~L.} \bibnamefont{Veatch}},
  \bibinfo{author}{\bibfnamefont{O.}~\bibnamefont{Soubias}},
  \bibinfo{author}{\bibfnamefont{S.~L.} \bibnamefont{Keller}},
  \bibnamefont{and}
  \bibinfo{author}{\bibfnamefont{K.}~\bibnamefont{Gawsrisch}},
  \bibinfo{journal}{PNAS} \textbf{\bibinfo{volume}{104}},
  \bibinfo{pages}{17650} (\bibinfo{year}{2007}).

\bibitem[{\citenamefont{Pokorny et~al.}(2006)\citenamefont{Pokorny, Yandek,
  Elegbede, Hinderliter, and Almeida}}]{pokorny}
\bibinfo{author}{\bibfnamefont{A.}~\bibnamefont{Pokorny}},
  \bibinfo{author}{\bibfnamefont{L.~E.} \bibnamefont{Yandek}},
  \bibinfo{author}{\bibfnamefont{A.~I.} \bibnamefont{Elegbede}},
  \bibinfo{author}{\bibfnamefont{A.}~\bibnamefont{Hinderliter}},
  \bibnamefont{and} \bibinfo{author}{\bibfnamefont{P.~F.~F.}
  \bibnamefont{Almeida}}, \bibinfo{journal}{Biophysical Journal}
  \textbf{\bibinfo{volume}{91}}, \bibinfo{pages}{2184} (\bibinfo{year}{2006}).

\bibitem[{\citenamefont{Bunge et~al.}(2008)\citenamefont{Bunge, M\"{u}ller,
  St\"{o}ckl, Herrmann, and Huster}}]{bunge}
\bibinfo{author}{\bibfnamefont{A.}~\bibnamefont{Bunge}},
  \bibinfo{author}{\bibfnamefont{P.~M.} \bibnamefont{M\"{u}ller}},
  \bibinfo{author}{\bibfnamefont{M.}~\bibnamefont{St\"{o}ckl}},
  \bibinfo{author}{\bibfnamefont{A.}~\bibnamefont{Herrmann}}, \bibnamefont{and}
  \bibinfo{author}{\bibfnamefont{D.}~\bibnamefont{Huster}},
  \bibinfo{journal}{Biophys. J.} \textbf{\bibinfo{volume}{94}},
  \bibinfo{pages}{2680} (\bibinfo{year}{2008}).

\bibitem[{\citenamefont{Halling et~al.}(2008)\citenamefont{Halling, Ramstedt,
  Nustr\"{o}m, Slotte, and Nyholm}}]{halling}
\bibinfo{author}{\bibfnamefont{K.~K.} \bibnamefont{Halling}},
  \bibinfo{author}{\bibfnamefont{B.}~\bibnamefont{Ramstedt}},
  \bibinfo{author}{\bibfnamefont{J.~H.} \bibnamefont{Nustr\"{o}m}},
  \bibinfo{author}{\bibfnamefont{J.~P.} \bibnamefont{Slotte}},
  \bibnamefont{and} \bibinfo{author}{\bibfnamefont{T.~K.~M.}
  \bibnamefont{Nyholm}}, \bibinfo{journal}{Biophysical Journal}
  \textbf{\bibinfo{volume}{95}}, \bibinfo{pages}{3861} (\bibinfo{year}{2008}).

\bibitem[{\citenamefont{de~Almeida et~al.}(2003)\citenamefont{de~Almeida,
  Feorov, and Prieto}}]{dealmeida}
\bibinfo{author}{\bibfnamefont{R.}~\bibnamefont{de~Almeida}},
  \bibinfo{author}{\bibfnamefont{A.}~\bibnamefont{Feorov}}, \bibnamefont{and}
  \bibinfo{author}{\bibfnamefont{M.}~\bibnamefont{Prieto}},
  \bibinfo{journal}{Biophys. J.} \textbf{\bibinfo{volume}{85}},
  \bibinfo{pages}{2406} (\bibinfo{year}{2003}).

\bibitem[{\citenamefont{Tsamaloukas et~al.}(2006)\citenamefont{Tsamaloukas,
  Szadkowska, and Heerklotz}}]{tsamaloukas}
\bibinfo{author}{\bibfnamefont{A.}~\bibnamefont{Tsamaloukas}},
  \bibinfo{author}{\bibfnamefont{H.}~\bibnamefont{Szadkowska}},
  \bibnamefont{and}
  \bibinfo{author}{\bibfnamefont{H.}~\bibnamefont{Heerklotz}},
  \bibinfo{journal}{Biophysical Journal} \textbf{\bibinfo{volume}{90}},
  \bibinfo{pages}{4479} (\bibinfo{year}{2006}).

\bibitem[{\citenamefont{Lingwood and Simons}(2010)}]{lingwood}
\bibinfo{author}{\bibfnamefont{D.}~\bibnamefont{Lingwood}} \bibnamefont{and}
  \bibinfo{author}{\bibfnamefont{K.}~\bibnamefont{Simons}},
  \bibinfo{journal}{Science} \textbf{\bibinfo{volume}{327}},
  \bibinfo{pages}{46} (\bibinfo{year}{2010}).

\bibitem[{\citenamefont{Simons and Ikonen}(1997)}]{simons}
\bibinfo{author}{\bibfnamefont{K.}~\bibnamefont{Simons}} \bibnamefont{and}
  \bibinfo{author}{\bibfnamefont{E.}~\bibnamefont{Ikonen}},
  \bibinfo{journal}{Nature} \textbf{\bibinfo{volume}{387}},
  \bibinfo{pages}{569} (\bibinfo{year}{1997}).

\bibitem[{\citenamefont{Edidin}(2003)}]{edidin}
\bibinfo{author}{\bibfnamefont{M.}~\bibnamefont{Edidin}},
  \bibinfo{journal}{Annu. Rev. Biophys. Biomol. Struct}
  \textbf{\bibinfo{volume}{32}}, \bibinfo{pages}{257} (\bibinfo{year}{2003}).

\bibitem[{\citenamefont{Jacobson et~al.}(2007)\citenamefont{Jacobson,
  Mouritsen, and Anderson}}]{jacobson}
\bibinfo{author}{\bibfnamefont{K.}~\bibnamefont{Jacobson}},
  \bibinfo{author}{\bibfnamefont{O.~G.} \bibnamefont{Mouritsen}},
  \bibnamefont{and} \bibinfo{author}{\bibfnamefont{R.~G.~W.}
  \bibnamefont{Anderson}}, \bibinfo{journal}{Nature Cell Biology}
  \textbf{\bibinfo{volume}{9}}, \bibinfo{pages}{7} (\bibinfo{year}{2007}).

\bibitem[{\citenamefont{Hancock}(2006)}]{hancock}
\bibinfo{author}{\bibfnamefont{J.}~\bibnamefont{Hancock}},
  \bibinfo{journal}{Nature Rev. Mol. Cell Biol.} \textbf{\bibinfo{volume}{7}},
  \bibinfo{pages}{456} (\bibinfo{year}{2006}).

\bibitem[{\citenamefont{Pike}(2003)}]{pike1}
\bibinfo{author}{\bibfnamefont{L.~J.} \bibnamefont{Pike}},
  \bibinfo{journal}{Journal of Lipid Research} \textbf{\bibinfo{volume}{44}},
  \bibinfo{pages}{655} (\bibinfo{year}{2003}).

\bibitem[{\citenamefont{Pike}(2004)}]{pike2}
\bibinfo{author}{\bibfnamefont{L.~J.} \bibnamefont{Pike}},
  \bibinfo{journal}{Biochemical Journal} \textbf{\bibinfo{volume}{378}},
  \bibinfo{pages}{281} (\bibinfo{year}{2004}).

\bibitem[{\citenamefont{Yeagle}(1993)}]{yeagle}
\bibinfo{author}{\bibfnamefont{P.}~\bibnamefont{Yeagle}},
  \emph{\bibinfo{title}{The Membranes of Cells -- 2nd. ed.}}
  (\bibinfo{publisher}{Academic Press, Inc.}, \bibinfo{year}{1993}).

\bibitem[{\citenamefont{Lantzsch et~al.}(1994)\citenamefont{Lantzsch, Binder,
  and Heerklotz}}]{lantzsch}
\bibinfo{author}{\bibfnamefont{G.}~\bibnamefont{Lantzsch}},
  \bibinfo{author}{\bibfnamefont{H.}~\bibnamefont{Binder}}, \bibnamefont{and}
  \bibinfo{author}{\bibfnamefont{H.}~\bibnamefont{Heerklotz}},
  \bibinfo{journal}{Journal of Fluorescence} \textbf{\bibinfo{volume}{4}},
  \bibinfo{pages}{339} (\bibinfo{year}{1994}).

\bibitem[{\citenamefont{Veatch and Keller}(2003)}]{veatch2}
\bibinfo{author}{\bibfnamefont{S.}~\bibnamefont{Veatch}} \bibnamefont{and}
  \bibinfo{author}{\bibfnamefont{S.}~\bibnamefont{Keller}},
  \bibinfo{journal}{Biophys. J.} \textbf{\bibinfo{volume}{85}},
  \bibinfo{pages}{3074} (\bibinfo{year}{2003}).

\bibitem[{\citenamefont{Korlach et~al.}(1999)\citenamefont{Korlach, Schwille,
  Webb, and Feigenson}}]{korlach}
\bibinfo{author}{\bibfnamefont{J.}~\bibnamefont{Korlach}},
  \bibinfo{author}{\bibfnamefont{P.}~\bibnamefont{Schwille}},
  \bibinfo{author}{\bibfnamefont{W.~W.} \bibnamefont{Webb}}, \bibnamefont{and}
  \bibinfo{author}{\bibfnamefont{G.~W.} \bibnamefont{Feigenson}},
  \bibinfo{journal}{PNAS} \textbf{\bibinfo{volume}{96}}, \bibinfo{pages}{8461}
  (\bibinfo{year}{1999}).

\bibitem[{\citenamefont{Vist and Davis}(1990)}]{vist}
\bibinfo{author}{\bibfnamefont{M.~R.} \bibnamefont{Vist}} \bibnamefont{and}
  \bibinfo{author}{\bibfnamefont{J.~H.} \bibnamefont{Davis}},
  \bibinfo{journal}{Biochemistry} \textbf{\bibinfo{volume}{29}},
  \bibinfo{pages}{451} (\bibinfo{year}{1990}).

\bibitem[{\citenamefont{McMullen and McElhaney}(1995)}]{mcmullen}
\bibinfo{author}{\bibfnamefont{T.~P.~W.} \bibnamefont{McMullen}}
  \bibnamefont{and} \bibinfo{author}{\bibfnamefont{R.~N.}
  \bibnamefont{McElhaney}}, \bibinfo{journal}{Biochimica et Biophysica Acta}
  \textbf{\bibinfo{volume}{1234}}, \bibinfo{pages}{90} (\bibinfo{year}{1995}).

\bibitem[{\citenamefont{Mateo et~al.}(1995)\citenamefont{Mateo, Acu{\~n}a, and
  Brochon}}]{mateo}
\bibinfo{author}{\bibfnamefont{C.~R.} \bibnamefont{Mateo}},
  \bibinfo{author}{\bibfnamefont{A.~U.} \bibnamefont{Acu{\~n}a}},
  \bibnamefont{and} \bibinfo{author}{\bibfnamefont{J.~C.}
  \bibnamefont{Brochon}}, \bibinfo{journal}{Biophys. J.}
  \textbf{\bibinfo{volume}{68}}, \bibinfo{pages}{978} (\bibinfo{year}{1995}).

\bibitem[{\citenamefont{Khelashvili et~al.}(2005)\citenamefont{Khelashvili,
  Pandit, and Scott}}]{scmft1}
\bibinfo{author}{\bibfnamefont{G.~A.} \bibnamefont{Khelashvili}},
  \bibinfo{author}{\bibfnamefont{S.~A.} \bibnamefont{Pandit}},
  \bibnamefont{and} \bibinfo{author}{\bibfnamefont{H.~L.} \bibnamefont{Scott}},
  \bibinfo{journal}{J. Chem. Phys.} \textbf{\bibinfo{volume}{123}},
  \bibinfo{pages}{024910} (\bibinfo{year}{2005}).

\bibitem[{\citenamefont{Pandit et~al.}(2007)\citenamefont{Pandit, Khelashvili,
  Jakobsson, Grama, and Scott}}]{scmft2}
\bibinfo{author}{\bibfnamefont{S.~A.} \bibnamefont{Pandit}},
  \bibinfo{author}{\bibfnamefont{G.~A.} \bibnamefont{Khelashvili}},
  \bibinfo{author}{\bibfnamefont{E.}~\bibnamefont{Jakobsson}},
  \bibinfo{author}{\bibfnamefont{A.}~\bibnamefont{Grama}}, \bibnamefont{and}
  \bibinfo{author}{\bibfnamefont{H.~L.} \bibnamefont{Scott}},
  \bibinfo{journal}{Biophys. J.} \textbf{\bibinfo{volume}{92}},
  \bibinfo{pages}{440} (\bibinfo{year}{2007}).

\bibitem[{\citenamefont{Pan et~al.}(2009)\citenamefont{Pan, Tristram-Nagle, and
  Nagle}}]{pan}
\bibinfo{author}{\bibfnamefont{J.}~\bibnamefont{Pan}},
  \bibinfo{author}{\bibfnamefont{S.}~\bibnamefont{Tristram-Nagle}},
  \bibnamefont{and} \bibinfo{author}{\bibfnamefont{J.~F.} \bibnamefont{Nagle}},
  \bibinfo{journal}{Physical Review E} \textbf{\bibinfo{volume}{80}},
  \bibinfo{pages}{021931} (\bibinfo{year}{2009}).

\bibitem[{\citenamefont{Ramstedt and Slotte}(2006)}]{ramstedt}
\bibinfo{author}{\bibfnamefont{B.}~\bibnamefont{Ramstedt}} \bibnamefont{and}
  \bibinfo{author}{\bibfnamefont{J.~P.} \bibnamefont{Slotte}},
  \bibinfo{journal}{Biochimica et Biophysica Acta}
  \textbf{\bibinfo{volume}{1758}}, \bibinfo{pages}{1945}
  (\bibinfo{year}{2006}).

\bibitem[{\citenamefont{Pandit et~al.}(2008)\citenamefont{Pandit, Chiu,
  Jakobsson, Grama, and Scott}}]{langmuir}
\bibinfo{author}{\bibfnamefont{S.~A.} \bibnamefont{Pandit}},
  \bibinfo{author}{\bibfnamefont{S.}~\bibnamefont{Chiu}},
  \bibinfo{author}{\bibfnamefont{E.}~\bibnamefont{Jakobsson}},
  \bibinfo{author}{\bibfnamefont{A.}~\bibnamefont{Grama}}, \bibnamefont{and}
  \bibinfo{author}{\bibfnamefont{H.~L.} \bibnamefont{Scott}},
  \bibinfo{journal}{Langmuir} \textbf{\bibinfo{volume}{24}},
  \bibinfo{pages}{6858} (\bibinfo{year}{2008}).

\bibitem[{\citenamefont{Baumgart et~al.}(2007)\citenamefont{Baumgart, Hunt,
  Farkas, Webb, and Feigenson}}]{baumgart}
\bibinfo{author}{\bibfnamefont{T.}~\bibnamefont{Baumgart}},
  \bibinfo{author}{\bibfnamefont{G.}~\bibnamefont{Hunt}},
  \bibinfo{author}{\bibfnamefont{E.~R.} \bibnamefont{Farkas}},
  \bibinfo{author}{\bibfnamefont{W.~W.} \bibnamefont{Webb}}, \bibnamefont{and}
  \bibinfo{author}{\bibfnamefont{G.~W.} \bibnamefont{Feigenson}},
  \bibinfo{journal}{Biochimica et Biophysica Acta}
  \textbf{\bibinfo{volume}{1768}}, \bibinfo{pages}{2182}
  (\bibinfo{year}{2007}).

\bibitem[{\citenamefont{Morales-Penningston
  et~al.}(2010)\citenamefont{Morales-Penningston, Wu, Farkas, Goh, Konykhina,
  Zheng, Webb, and Feigenson}}]{morales}
\bibinfo{author}{\bibfnamefont{N.~F.} \bibnamefont{Morales-Penningston}},
  \bibinfo{author}{\bibfnamefont{J.}~\bibnamefont{Wu}},
  \bibinfo{author}{\bibfnamefont{E.~R.} \bibnamefont{Farkas}},
  \bibinfo{author}{\bibfnamefont{S.~L.} \bibnamefont{Goh}},
  \bibinfo{author}{\bibfnamefont{T.~M.} \bibnamefont{Konykhina}},
  \bibinfo{author}{\bibfnamefont{J.~Y.} \bibnamefont{Zheng}},
  \bibinfo{author}{\bibfnamefont{W.~W.} \bibnamefont{Webb}}, \bibnamefont{and}
  \bibinfo{author}{\bibfnamefont{G.~W.} \bibnamefont{Feigenson}},
  \bibinfo{journal}{BBA - Biomembranes}  (\bibinfo{year}{2010}).

\bibitem[{\citenamefont{Marsh}(2009)}]{marsh}
\bibinfo{author}{\bibfnamefont{D.}~\bibnamefont{Marsh}},
  \bibinfo{journal}{Biochimica et Biophysica Acta}
  \textbf{\bibinfo{volume}{1788}}, \bibinfo{pages}{2114}
  (\bibinfo{year}{2009}).

\bibitem[{\citenamefont{Lichtenberg et~al.}(2005)\citenamefont{Lichtenberg,
  Go{\~n}i, and Heerklotz}}]{lichtenberg}
\bibinfo{author}{\bibfnamefont{D.}~\bibnamefont{Lichtenberg}},
  \bibinfo{author}{\bibfnamefont{F.~M.} \bibnamefont{Go{\~n}i}},
  \bibnamefont{and}
  \bibinfo{author}{\bibfnamefont{H.}~\bibnamefont{Heerklotz}},
  \bibinfo{journal}{Trends in Biochemical Sciences}
  \textbf{\bibinfo{volume}{8}}, \bibinfo{pages}{430} (\bibinfo{year}{2005}).

\bibitem[{\citenamefont{Putzel and Schick}(2008)}]{putzel1}
\bibinfo{author}{\bibfnamefont{G.~G.} \bibnamefont{Putzel}} \bibnamefont{and}
  \bibinfo{author}{\bibfnamefont{M.}~\bibnamefont{Schick}},
  \bibinfo{journal}{Biophysical Journal} \textbf{\bibinfo{volume}{95}},
  \bibinfo{pages}{4756} (\bibinfo{year}{2008}).

\bibitem[{\citenamefont{Putzel and Schick}(2009)}]{putzel2}
\bibinfo{author}{\bibfnamefont{G.~G.} \bibnamefont{Putzel}} \bibnamefont{and}
  \bibinfo{author}{\bibfnamefont{M.}~\bibnamefont{Schick}},
  \bibinfo{journal}{Biophysical Journal} \textbf{\bibinfo{volume}{96}},
  \bibinfo{pages}{4935} (\bibinfo{year}{2009}).

\bibitem[{\citenamefont{Elliott et~al.}(2006)\citenamefont{Elliott, Szleifer,
  and Schick}}]{elliott}
\bibinfo{author}{\bibfnamefont{R.}~\bibnamefont{Elliott}},
  \bibinfo{author}{\bibfnamefont{I.}~\bibnamefont{Szleifer}}, \bibnamefont{and}
  \bibinfo{author}{\bibfnamefont{M.}~\bibnamefont{Schick}},
  \bibinfo{journal}{Phys. Rev. Lett.} \textbf{\bibinfo{volume}{96}},
  \bibinfo{pages}{098101} (\bibinfo{year}{2006}).

\bibitem[{\citenamefont{Elliott et~al.}(2005)\citenamefont{Elliott, Katsov,
  Schick, and Szleifer}}]{elliot2}
\bibinfo{author}{\bibfnamefont{R.}~\bibnamefont{Elliott}},
  \bibinfo{author}{\bibfnamefont{K.}~\bibnamefont{Katsov}},
  \bibinfo{author}{\bibfnamefont{M.}~\bibnamefont{Schick}}, \bibnamefont{and}
  \bibinfo{author}{\bibfnamefont{I.}~\bibnamefont{Szleifer}},
  \bibinfo{journal}{The Journal of Chemical Physics}
  \textbf{\bibinfo{volume}{122}}, \bibinfo{pages}{044904}
  (\bibinfo{year}{2005}).

\bibitem[{\citenamefont{Tumaneng et~al.}(2010)\citenamefont{Tumaneng, Pandit,
  Zhao, and Scott}}]{jcp}
\bibinfo{author}{\bibfnamefont{P.~W.} \bibnamefont{Tumaneng}},
  \bibinfo{author}{\bibfnamefont{S.~A.} \bibnamefont{Pandit}},
  \bibinfo{author}{\bibfnamefont{G.}~\bibnamefont{Zhao}}, \bibnamefont{and}
  \bibinfo{author}{\bibfnamefont{H.~L.} \bibnamefont{Scott}},
  \bibinfo{journal}{The Journal of Chemical Physics}
  \textbf{\bibinfo{volume}{132}}, \bibinfo{pages}{065104}
  (\bibinfo{year}{2010}).

\bibitem[{\citenamefont{Pandit and Scott}(2009)}]{scmft3}
\bibinfo{author}{\bibfnamefont{S.~A.} \bibnamefont{Pandit}} \bibnamefont{and}
  \bibinfo{author}{\bibfnamefont{H.~L.} \bibnamefont{Scott}},
  \bibinfo{journal}{Biochim. Biophys. Acta} \textbf{\bibinfo{volume}{1788}},
  \bibinfo{pages}{136} (\bibinfo{year}{2009}).

\bibitem[{\citenamefont{Shi and Voth}(2005)}]{shi}
\bibinfo{author}{\bibfnamefont{Q.}~\bibnamefont{Shi}} \bibnamefont{and}
  \bibinfo{author}{\bibfnamefont{G.~A.} \bibnamefont{Voth}},
  \bibinfo{journal}{Biophys. J.} \textbf{\bibinfo{volume}{89}},
  \bibinfo{pages}{2385} (\bibinfo{year}{2005}).

\bibitem[{\citenamefont{Idema et~al.}(2009)\citenamefont{Idema, van Leeuwen,
  and Storm}}]{idema}
\bibinfo{author}{\bibfnamefont{T.}~\bibnamefont{Idema}},
  \bibinfo{author}{\bibfnamefont{J.~M.~J.} \bibnamefont{van Leeuwen}},
  \bibnamefont{and} \bibinfo{author}{\bibfnamefont{C.}~\bibnamefont{Storm}},
  \bibinfo{journal}{Phys. Rev. E} \textbf{\bibinfo{volume}{80}},
  \bibinfo{pages}{041924} (\bibinfo{year}{2009}).

\bibitem[{\citenamefont{Fattal and Ben-Shaul}(1994)}]{fattal}
\bibinfo{author}{\bibfnamefont{D.~R.} \bibnamefont{Fattal}} \bibnamefont{and}
  \bibinfo{author}{\bibfnamefont{A.}~\bibnamefont{Ben-Shaul}},
  \bibinfo{journal}{Biophysical Journal} \textbf{\bibinfo{volume}{67}},
  \bibinfo{pages}{983} (\bibinfo{year}{1994}).

\bibitem[{\citenamefont{Ben-Shaul et~al.}(1985)\citenamefont{Ben-Shaul,
  Szleifer, and Gelbart}}]{ben-shaul}
\bibinfo{author}{\bibfnamefont{A.}~\bibnamefont{Ben-Shaul}},
  \bibinfo{author}{\bibfnamefont{I.}~\bibnamefont{Szleifer}}, \bibnamefont{and}
  \bibinfo{author}{\bibfnamefont{W.~M.} \bibnamefont{Gelbart}},
  \bibinfo{journal}{J. Chem. Phys.} \textbf{\bibinfo{volume}{83}},
  \bibinfo{pages}{3597} (\bibinfo{year}{1985}).

\bibitem[{\citenamefont{Gruen}(1985)}]{gruen}
\bibinfo{author}{\bibfnamefont{D.~W.~R.} \bibnamefont{Gruen}},
  \bibinfo{journal}{J. Phys. Chem} \textbf{\bibinfo{volume}{89}},
  \bibinfo{pages}{146} (\bibinfo{year}{1985}).

\bibitem[{\citenamefont{M\"{u}ller et~al.}(2006)\citenamefont{M\"{u}ller,
  Katsov, and Schick}}]{muller}
\bibinfo{author}{\bibfnamefont{M.}~\bibnamefont{M\"{u}ller}},
  \bibinfo{author}{\bibfnamefont{K.}~\bibnamefont{Katsov}}, \bibnamefont{and}
  \bibinfo{author}{\bibfnamefont{M.}~\bibnamefont{Schick}},
  \bibinfo{journal}{Physics Reports} \textbf{\bibinfo{volume}{434}},
  \bibinfo{pages}{113} (\bibinfo{year}{2006}).

\bibitem[{\citenamefont{McConnell}(2009)}]{mcconnell}
\bibinfo{author}{\bibfnamefont{H.}~\bibnamefont{McConnell}},
  \bibinfo{journal}{The Journal of Chemical Physics}
  \textbf{\bibinfo{volume}{130}}, \bibinfo{pages}{165103}
  (\bibinfo{year}{2009}).

\bibitem[{\citenamefont{Risselada and Marrink}(2008)}]{risselada}
\bibinfo{author}{\bibfnamefont{H.~J.} \bibnamefont{Risselada}}
  \bibnamefont{and} \bibinfo{author}{\bibfnamefont{S.~J.}
  \bibnamefont{Marrink}}, \bibinfo{journal}{PNAS}
  \textbf{\bibinfo{volume}{105}}, \bibinfo{pages}{17367}
  (\bibinfo{year}{2008}).

\bibitem[{\citenamefont{Maulik and Shipley}(1996)}]{maulik}
\bibinfo{author}{\bibfnamefont{P.~R.} \bibnamefont{Maulik}} \bibnamefont{and}
  \bibinfo{author}{\bibfnamefont{G.~G.} \bibnamefont{Shipley}},
  \bibinfo{journal}{Biophys. J.} \textbf{\bibinfo{volume}{70}},
  \bibinfo{pages}{2256} (\bibinfo{year}{1996}).

\bibitem[{\citenamefont{Chaiken and Lubensky}(1995)}]{chaiken}
\bibinfo{author}{\bibfnamefont{P.~M.} \bibnamefont{Chaiken}} \bibnamefont{and}
  \bibinfo{author}{\bibfnamefont{T.~C.} \bibnamefont{Lubensky}},
  \emph{\bibinfo{title}{Principles of Condensed Matter Physics}}
  (\bibinfo{publisher}{Cambridge University Press}, \bibinfo{year}{1995}).

\bibitem[{\citenamefont{Mar\u{c}elja}(1974)}]{marcelja}
\bibinfo{author}{\bibfnamefont{S.}~\bibnamefont{Mar\u{c}elja}},
  \bibinfo{journal}{Biochim. Biophys. Acta} \textbf{\bibinfo{volume}{367}},
  \bibinfo{pages}{156} (\bibinfo{year}{1974}).

\bibitem[{\citenamefont{Berendsen et~al.}(1995)\citenamefont{Berendsen, van~der
  Spoel, and van Drunen}}]{berendsen}
\bibinfo{author}{\bibfnamefont{H.}~\bibnamefont{Berendsen}},
  \bibinfo{author}{\bibfnamefont{D.}~\bibnamefont{van~der Spoel}},
  \bibnamefont{and} \bibinfo{author}{\bibfnamefont{R.}~\bibnamefont{van
  Drunen}}, \bibinfo{journal}{Comp. Phys. Comm} \textbf{\bibinfo{volume}{91}},
  \bibinfo{pages}{43} (\bibinfo{year}{1995}).

\bibitem[{\citenamefont{Lindahl et~al.}(2001)\citenamefont{Lindahl, Hess, and
  van~der Spoel}}]{lindahl}
\bibinfo{author}{\bibfnamefont{E.}~\bibnamefont{Lindahl}},
  \bibinfo{author}{\bibfnamefont{B.}~\bibnamefont{Hess}}, \bibnamefont{and}
  \bibinfo{author}{\bibfnamefont{D.}~\bibnamefont{van~der Spoel}},
  \bibinfo{journal}{J. Mol. Mod.} \textbf{\bibinfo{volume}{7}},
  \bibinfo{pages}{306} (\bibinfo{year}{2001}).

\bibitem[{\citenamefont{Hess et~al.}(1997)\citenamefont{Hess, Bekker,
  Berendsen, and Fraaige}}]{hess}
\bibinfo{author}{\bibfnamefont{B.}~\bibnamefont{Hess}},
  \bibinfo{author}{\bibfnamefont{H.}~\bibnamefont{Bekker}},
  \bibinfo{author}{\bibfnamefont{H.~J.~C.} \bibnamefont{Berendsen}},
  \bibnamefont{and} \bibinfo{author}{\bibfnamefont{J.~G. E.~M.}
  \bibnamefont{Fraaige}}, \bibinfo{journal}{J. Comp. Chem.}
  \textbf{\bibinfo{volume}{18}} (\bibinfo{year}{1997}).

\bibitem[{\citenamefont{Essmann et~al.}(1995)\citenamefont{Essmann, Perera,
  Berkowitz, Darden, Lee, and Pedersen}}]{essmann}
\bibinfo{author}{\bibfnamefont{U.}~\bibnamefont{Essmann}},
  \bibinfo{author}{\bibfnamefont{L.}~\bibnamefont{Perera}},
  \bibinfo{author}{\bibfnamefont{M.~L.} \bibnamefont{Berkowitz}},
  \bibinfo{author}{\bibfnamefont{T.}~\bibnamefont{Darden}},
  \bibinfo{author}{\bibfnamefont{H.}~\bibnamefont{Lee}}, \bibnamefont{and}
  \bibinfo{author}{\bibfnamefont{L.~G.} \bibnamefont{Pedersen}},
  \bibinfo{journal}{J. Chem. Phys.} \textbf{\bibinfo{volume}{103}},
  \bibinfo{pages}{8577} (\bibinfo{year}{1995}).

\bibitem[{\citenamefont{Nose and Klein}(1983)}]{nose}
\bibinfo{author}{\bibfnamefont{S.}~\bibnamefont{Nose}} \bibnamefont{and}
  \bibinfo{author}{\bibfnamefont{M.~L.} \bibnamefont{Klein}},
  \bibinfo{journal}{Mol. Phys.} \textbf{\bibinfo{volume}{50}},
  \bibinfo{pages}{1055} (\bibinfo{year}{1983}).

\bibitem[{\citenamefont{Parrinello and Rahman}(1981)}]{parrinello}
\bibinfo{author}{\bibfnamefont{M.}~\bibnamefont{Parrinello}} \bibnamefont{and}
  \bibinfo{author}{\bibfnamefont{A.}~\bibnamefont{Rahman}},
  \bibinfo{journal}{J. Appl. Phys} \textbf{\bibinfo{volume}{52}},
  \bibinfo{pages}{7182} (\bibinfo{year}{1981}).

\bibitem[{\citenamefont{Peng et~al.}(2000)\citenamefont{Peng, Qiu, Ginzburg,
  Jasnow, and Balazs}}]{peng}
\bibinfo{author}{\bibfnamefont{G.}~\bibnamefont{Peng}},
  \bibinfo{author}{\bibfnamefont{F.}~\bibnamefont{Qiu}},
  \bibinfo{author}{\bibfnamefont{V.}~\bibnamefont{Ginzburg}},
  \bibinfo{author}{\bibfnamefont{D.}~\bibnamefont{Jasnow}}, \bibnamefont{and}
  \bibinfo{author}{\bibfnamefont{A.}~\bibnamefont{Balazs}},
  \bibinfo{journal}{Science} \textbf{\bibinfo{volume}{288}},
  \bibinfo{pages}{1802} (\bibinfo{year}{2000}).

\end{thebibliography}
\end{document}